\documentclass[referee]{raa}            % referee version: for submission

\usepackage{graphicx,times}             %for PS/EPS graphics inclusion, new
\usepackage{natbib}
\usepackage{amssymb,amsmath}
\bibpunct{(}{)}{;}{a}{}{,}

\def\dsm{$M_\odot$}

\usepackage[a4paper=true,dvipdfm=true,pagebackref=true]{hyperref}
\hypersetup{colorlinks = true, linkcolor = green, anchorcolor = red, citecolor = blue, filecolor = red, pagecolor = red, urlcolor = red}

\begin{document}

   \title{A database of spectral energy distributions of progenitors of core-collapse supernovae\footnote{The database and SED fitting code are available at GitHub.}
%\footnotetext{$*$ Supported by the National Natural Science Foundation of China.}
}
%   \subtitle{I. Place Your Subtitle Here}

   \volnopage{Vol.0 (20xx) No.0, 000--000}      %%preserved for Editor. DOn't remove!
   \setcounter{page}{1}          %%starting page, preserved for Editor. DOn't remove!

   \author{Zhong-Mu Li
      \inst{1}
      \and Cai-Yan Mao
      \inst{1}
   }

   \institute{Institute for Astronomy and History of Science and
Technology, Dali University, Dali 671003, China; {\it zhongmuli@126.com}\\
 {\small Received~~20xx month day; accepted~~20xx~~month day}}

\abstract{This paper presents a database of the spectroscopic- and photometric- spectral energy distributions (spec-SEDs and phot-SEDs) of the progenitors of core-collapse supernovae (CCSNe). Both binary- and single-star progenitors are included in the database. The database covers the initial metallicity ($Z$) range of 0.0001--0.03, mass range of 8--25 \dsm{}, binary mass ratio range of 0--1, and orbital period range of 0.1--10000\,days. The low-resolution spec-SEDs and phot-SEDs of single- and binary-star CCSN progenitors are included in the database. These data can be used for studying the basic parameters, e.g., metallicity, age, initial and final masses of CCSN progenitors. It can also be used for studying the effects of different factors on the determination of parameters of CCSN progenitors. When the database is used for fitting the SEDs of binary-star CCSN progenitors, it is strongly suggested to determine the metallicity and orbital period in advance, while it is not necessary for single-star progenitors.
}

   \authorrunning{Li \& Mao}            %author_head in even pages
   \titlerunning{SEDs of CCSN Progenitors}  % title_head in odd pages

   \maketitle
%% The author head (on even pages) and the title head (on odd pages) will be
%% automatically extracted from \author{} and \title{}. Whenever the title is too long,
%% you will be asked to supply a shorter one by inserting either \authorrunning{} or
%% \titlerunning{} before \maketitle. Anyway, you can specify your own heads.
%%
%%
%% Note: In the following text body of your manuscript, please note several differences from
%%       other major journals:
%% (1) \subsection{Please Capitalize the First Letter of Each Notional Word in Subsection Title}
%% (2) Please Capitalize the First Letter of Each Notional Word in all tables' captions

%
%________________________________________________ sections below
%

%% Authors can give a citation as 'Michel et al. 1992'.
%% You may also use \cite, \citep and \citet for citation, and use Table~1 or Figure~1
%% and so forth. Using \ref and \label for cross-references of Tables/Figures
%% is a good way in adjusting/adding/removing text, tables or figures.

\section{Introduction}           %% first-level sections will be auto-capitalized
\label{sect:intro}

As the main kind of supernovae (SNe), core-collapse supernova (CCSN) is the explosion that attend the death of massive stars.
CCSN is a singularly important phenomenon in the universe for two main reasons.
First, CCSNe are principal drivers of cosmic chemical evolution. Most heavy elements heavier than hydrogen (H) and helium (He),
except those around the iron peak, were synthesized by CCSNe.
Second, they are possibly related to the rapid neutron capture process (r-process), which produced many of the extremely heavy elements above atomic mass of approximately 70.
CCSNe have observed kinetic energies of typically $\sim$10$^{51}$ ergs, and their integrated luminosities are usually 1-–10\% of this value \citep{2009ARAA..47...63S}.
The explosion of CCSNe has been a perennial challenge in theoretical astrophysics for decades. So does the progenitor of CCSNe.

The progenitor of CCSNe is fundamental for understanding CCSNe, in particular, for their explosions.
However, it is still far from well understanding the formation and properties of CCSN progenitors.
It is well-known that the minimum initial mass that can produce a CCSN is about 8 \dsm{}, according to the direct detections of red supergiant progenitors and the most massive white dwarf progenitors. The maximum initial mass is less than about 25 \dsm{}, because the majority of massive stars above 20\dsm{} may collapse quietly to black holes and that the explosions remain undetected. The progenitors of CCSNe have been widely studied, but the results are actually model dependent \citep{2009ARAA..47...63S}.
The common image to form a CCSN progenitor is as follows.
The H in stellar cores converts to He in stellar evolution.
If a star is sufficiently massive, heavier elements such as carbon, oxygen, nickel, nitrogen, magnesium, silicon and iron (C, O, Ne, N,
Mg, Si and Fe) are subsequently produced in nuclear synthesis reactions. For stars more massive than 8 \dsm{}, either an O-Ne-Mg core \citep{2008ApJ...675..614P}
or a Fe core \citep{2002RvMP...74.1015W} will form eventually and would cause an SN explosion \citep{lisakov_tel-02018238}.

Most studies of CCSN progenitors so far take single-star models. For example, the nearest progenitor, SN1987A, which is in the Large Magellanic Cloud (LMC), is shown to be evolved  from a single star with initial mass in the region of 14–-20 \dsm{}. Using detailed stellar evolutionary codes such as MESA \citep{2011ApJS..192....3P}, many works evolved massive star models from the main sequence until core collapse (e.g., \citealt{2018MNRAS.473.3863L}). These works put forward a lot of the studies of CCSN progenitors.
However, single-star models ignore the observed fact that a lot of stars are in binaries.
In fact, many of the supernova progenitors are possibly binaries. For example, CCSN SN1993J needs a binary of 15+14 \dsm{} with an initial
orbital period of 5.8 years to explain the observed UBVRCIc SED \citep{1994PASP..106.1025H}.
A potential surviving companion of the type Ia supernova Tycho Brahe’s 1572, was found around the position of the explosion \citep{2004Natur.431.1069R}.

According to stellar evolution theory, binary stars evolve in a substantially different way from single stars if their components are not too far from each other.
Binary interactions are therefore very important for stellar evolution \citep{1992ApJ...391..246P}.
Binary evolution affects both the population synthesis of a large amount of stars \citep{2002ApJ...565.1107P,2008ApJS..174..223B, 2007MNRAS.380.1098H, 2008ApJ...685..225L, 2006MNRAS.370.1181Z,2017PASA...34...58E,2020MNRAS.494L..53F} and the detailed models of a small number of stars. If binary evolution is taken instead of single-star evolution, CCSN progenitors could differ a lot in mass, age, radius, and composition \citep{2020MNRAS.494L..53F, 2020arXiv200207230Z}, which would impact the resulting SN and its remnant obviously.

There has been a long history for the study of binary-star progenitors.
For example, \cite{2008MNRAS.384.1109E} investigated the effect of massive binaries on stellar populations and SN
progenitors using a detailed stellar evolution code \citep{2004MNRAS.353...87E} and the single-star evolution equation of \cite{2000MNRAS.315..543H}.
Their theoretical predictions from binary-star models agree with the observed ratios of the Type Ib/c SN rate to the
Type II SN rate at different metallicities, but the single-star models predict a lower relative rate for Type Ib/c SNe than the observation. This implies that many CCSNe stem from binary-star progenitors. However, their results cannot be used for studying the SEDs of CCSNe progenitors because SEDs were not calculated and the binary parameter-space resolution of that work is rather low.
\cite{2008ApJ...685.1103W} studied the the heaviest models which do not encounter CCSN, and obtain some similar results with previous results (e.g., \citealt{2007Natur.450..390W,2008ApJ...673.1014U}).
The properties of Type Ib and IIb SN progenitors that are produced by stable mass transfer in binary
systems were explored using the MESA stellar evolution code \citep{2011ApJS..192....3P,2017ApJ...840...10Y} from an initial primary mass in the range of 10–-18 \dsm{} at solar and LMC metallicities. However, only two metallicities were considered and it is not enough for many studies of CCSN progenitors, in particular the SED studies.

Besides binarity, stellar rotation and magnetic fields were also not considered in most CCSN progenitor studies,
although they have some effects on the formation and properties of progenitors (e.g., \citealt{2000ApJ...528..368H,2007AA...464L..11M,2005NatPh...1..147W,2012ARAA..50..107L}). In fact, there is still a long way to go, because the effect of stellar rotation and magnetic field remains very uncertain (e.g., \citealt{2020MNRAS.494.4665P}).

In the studies of CCSNe, there have been a few good algorithms, e.g., Supernova Identification (SNID). Such codes can be used to identify the type of an SN spectrum and to determine its redshift and age \citep{2007ApJ...666.1024B}. However, there is no comprehensive SED database to determine the properties of different kinds of CCSN progenitors yet.
This hampers many studies, e.g., the identification of the CCSN progenitor on pre-explosion images.
This work therefore aims to build a database of the photometric- and spectroscopic- spectral energy distributions (phot- and spec-SEDs) of CCSN progenitors. Both photo-SEDs and spec-SEDs are concerned here because they are the main approaches to estimate the properties of CCSN progenitors. This is the first attempt to give the predicted SEDs of CCSN progenitors.

The structure of paper is as follows: in section 2, we introduce the parameter ranges of stars and the calculation of stellar evolution.
Then in section 3, we present the phot- and spec-SEDs of CCSN progenitors.
Next, in section 4, we apply the database to some mock progenitors with phot-SEDs.
Finally, we conclude and discuss this work in section 5.

%__________________________________________________________________

\section{Stellar parameter ranges and evolution computations}
\subsection{Parameter ranges}
This work aims to supply an SED database with a large parameter coverage and reasonable resolution, so the parameter ranges are wider than most previous works.
In detail, stellar metallicity ($Z$) covers a range of 0.0001--0.03. Stars from metal-poor to metal-rich kinds are included.
The zero-age main-sequence mass range of single stars is set to 8--25 \dsm{}, because most CCSN progenitors have main-sequence masses in this range.
This range is similar to some theoretical studies (e.g., \citealt{2018EPJP..133..388S}),
and larger than some observational results (8.5–-16.5 \dsm{}, e.g., \citealt{2009ARAA..47...63S} and references therein).
The same range is set to the total mass ($M_{\rm 1}$+$M_{\rm 2}$) of two binary components. The range of the mass ratio of secondary to primary of binaries, $q$, is set to 0--1.
The orbital period ($P$) of a binary changes from 0.1 to 10$^{4}$\,days. In fact, within the current age of the universe, the evolution of binaries with periods longer than 10$^{4}$\,days is similar to the counterparts with a period of 10$^{4}$\,days. The intervals of $M_{\rm 1}$, $q$ and log$P$ of main-sequence stars are set to 0.1, 0.1, 0.5, respectively. Two values (0.3 and 0.7) are chosen for the eccentricity ($e$) of binary stars, as previous studies (e.g., \citealt{2002MNRAS.329..897H}) have shown that $e$ affects the final results somewhat slightly.
Although the evolution of massive stars is sensitive to metallicity, binarity, rotation, and possibly magnetic fields, rotation and magnetic field are not taken into account in this work because of their huge uncertainties.
Table 1 lists the parameter ranges and steps of zero-age main sequence stars, which are taken by this work.

\begin{table*} %Table 1
 \caption{Parameter ranges and steps of zero-age main sequence stars of CCSN progenitors. $q$, $P$ and $e$ are for binary stars only.
 Stellar mass $M_{\rm 0}$ mean the total main-sequence mass of a single or binary star. For binaries, the masses of primary and secondary components are calculated by $M_{\rm 1}$ = $M_{\rm 0}/(1 + q)$ and $M_{\rm 2}$ = $q \times M_{\rm 1}$.}
 \begin{center}
  \begin{tabular}{cccccccccc}
   \hline
      Parameter     &Range               &Step          &Unit     &Note              \\
   \hline
      $Z$           &0.0001--0.03        &8 values      &         &single star and binary star\\
      $M_{\rm0}$   &8.0--25.0           &0.1           &\dsm{}    &single star and binary star\\
      $q$           &0--1.0              &0.1           &         &binary star           \\
      log($P$)      &-1.0--4.0           &0.5           &days     &binary star           \\
      $e$           &0.3--0.7            &0.4           &         &binary star           \\
   \hline
  \end{tabular}
 \end{center}
\end{table*}

Note that the mass range of this work is similar to most previous studies and findings.
For example, \cite{2009ARAA..47...63S} investigates a mass range of 8--25 \dsm{}.
\cite{2018MNRAS.473.3863L} takes 12, 25 and 27 \dsm{} in their work.
\cite{2010MNRAS.408..827D} performed some radiation-hydrodynamic simulations and indicate that the progenitor main-sequence masses inferior
to $\sim$ 20$\,M_{\odot}$, and the range of 25-–30 \dsm{} is not supported by the narrow width of OI 6303--6363 ${\rm \AA}$ in Type II-P SNe
with nebular spectra.
\cite{2012ARAA..50..107L} gave the likely minimum initial mass range of massive stars at solar metallicity as 8--12 \dsm{}, according to \cite{2008ApJ...675..614P}.
In close binaries, this limit depends on other initial system parameters such as metallicity and orbital period.
The mass limit at solar metallicity can be as high as 15 \dsm{} \citep{2001AA...369..939W}.

\subsection{Calculation of stellar evolution and progenitor properties}
This work models the parameters of CCSN progenitors with a reasonable resolution, via a rapid population-synthesis
code, BSE \citep{2000MNRAS.315..543H,2002MNRAS.329..897H}. It takes some fitting formulae based on the reliable stellar evolutionary tracks of \cite{1998MNRAS.298..525P}.
In addition to all aspects of single-star evolution, binary interactions including mass transfer, mass accretion, common-envelope evolution, collisions, supernova kicks and angular momentum loss mechanisms have been taken into account by this code, and the calculation result is similar to some detailed stellar evolution codes \citep{2012ARAA..50..107L}.
This code is fast for modeling the population of a large number of single or binary stars. It is widely used in many stellar population synthesis studies, e.g., \cite{2006MNRAS.370.1181Z,2007MNRAS.380.1098H,2012ApJ...761L..22L,2013ApJ...776...37L,2016ApJS..225....7L,2018NewA...64...61L,2018ApJ...859...36L}.
It has also been used by some previous works, e.g., \cite{2004MNRAS.353...87E} and \cite{2008MNRAS.384.1109E}, to reproduce the observed trends such as the distribution of well-studied SN progenitors in the metallicity versus initial mass plane, and the ratio of the Type Ib/c SN rate to the Type II SN rate.
The code makes it possible to cover large ranges of parameters in the studies of CCSN progenitors.
Although there are small uncertainties {($\leq$ 5 per cent) in the luminosity, radius and core mass compared to detailed stellar evolution codes,
the accuracy is enough for most SED studies of CCSN progenitors, as the uncertainties of phot-SEDs of distant CCSN progenitors are usually larger.

When evolving stars, some default values of BSE code are taken for the input parameters, because they have been checked by the developer and widely used in different works.
They are listed in Table 2 to help the readers to understand the physical processes in the evolution of CCSN progenitors. Note that there is an important difference between single stars and close binary components. Close binary components undergo mass transfer following Roche lobe over-follow but single stars do not. Mass transfer can occur between two binary components including different types. White dwarfs are the only degenerate objects able to fill their Roche lobes for a significant amount of time without breaking up. Thus dynamical mass transfer from a white dwarf can occur in binary evolution. Mass accretion on to degenerate objects is important both during Roche lobe overflow and when material is accreted from the wind of the companion. Accretion is assumed to be restricted by the Eddington limit. Two binary components can merge to a single remnant in some cases. Besides nondegenerate stars and white dwarfs, neutron stars and black holes can also merge. This will increase the mass and possibly change the type of the remnants. The angular momentum loss that is caused by both gravitational radiation and magnetic braking is considered by the BSE code. One can read \cite{1997MNRAS.291..732T, 1998MNRAS.298..525P, 2002MNRAS.329..897H} for more details about the treatment of stellar evolutionary processes.
\begin{table*} %Table 2
 \caption{Input parameters used for stellar evolution.}
 \begin{center}
  \begin{tabular}{cccccccccc}
   \hline
      Parameter or process                                   &Symbol          &Value               &Note              \\
   \hline
      Reimers mass-loss coefficent                           &$\eta$          &0.5                 &                  \\
      helium star mass loss factor                           &                &1.0                 &                  \\
      common-envelope efficiency parameter                   &$\alpha$        &3.0                 &                  \\
      binding energy factor for CE                           &$\lambda$       &0.5                 &                  \\
      spin-energy correction in common-envelope              &ceflag          &0                   &off               \\
      tidal circularisation                                  &tflag           &1                   &on                \\
      using modified-Mestel cooling for WDs                  &wdflag          &1                   &on                \\
      allowing velocity kick at BH formation                 &bhflag          &0                   &no                \\
      taking NS/BH mass from Belczynski et al. (2002)        &nsflag          &1                   &yes               \\
      maximum NS mass                                        &mxns            &3.0                 &in \dsm{}         \\
      dispersion in the Maxwellian for SN kick speed         &$\sigma$        &190.0               &in km/s            \\
      wind velocity factor                                   &$\beta$         &0.125       &$\propto$ $v_{\rm wind}^2$ \\
      the wind accretion efficiency factor                   &xi              &1.0                 &                  \\
      Bondi-Hoyle wind accretion factor                      &acc2            &1.5                 &                  \\
      fraction of accreted matter retained in nova eruption  &$\epsilon_{\rm nov}$  &0.001         &                  \\
      Eddington limit factor for mass transfer               &eddfac          &10.0                &                  \\
      angular momentum factor for mass lost during Roche     &$\gamma$        &-1.0                &                  \\
   \hline
  \end{tabular}
 \end{center}
\end{table*}

The progenitors of CCSNe are massive stars that evolve very fast. An C-O core or Fe core is finally formed and its mass grows with stellar evolution, up to the effective Chandrasekhar mass ($\geq$ about 1.26 \dsm{}). Once the core attains this critical mass, unstable gravitational collapse and explosion ensues.
This work gives the properties of CCSN progenitors at the moment of the explosion.
The CCSN progenitors are found by comparing the change of star type and the mass of stellar core, which is similar to the method of \cite{2008MNRAS.384.1109E}.
The phot- and spec-SEDs, age, mass, effective temperature, luminosity, gravitational acceleration, and radius are given for each progenitor.
If the progenitor is a binary, orbital period and eccentricity are also given.

A standard and widely used stellar spectral library, BaSeL 3.1 \citep{1997AAS..125..229L,1998AAS..130...65L}, is used for transforming the stellar evolutionary parameters to spec-SEDs when calculating the SEDs of CCSN progenitors. The SEDs cover a large wavelength from ultraviolet (UV) to medium infrared (MIR), which is suitable for most multi-band studies. This library is a comprehensive hybrid library of synthetic stellar spectra based on three original grids of model atmosphere spectra. It covers the largest possible ranges in stellar parameters (T$_{\rm eff}$, log$g$, and [M/H]) and provides flux spectra with useful resolution on an uniform grid of wavelengths. The standard library has been calibrated and its consistency has been tested carefully. In particular, the library spectra was conformed to the empirical color temperature relations, successfully.
After the calculation of spec-SEDs, the phot-SEDs are calculated from spec-SEDs, by taking the AB photometry system. All magnitudes of progenitors are calibrated using the data of Vega.

\section{Properties of CCSN progenitors}
The properties are given with the same format for single- and binary-star progenitors.
A single-star progenitor is regarded as a binary-star progenitor with a zero-mass component.
The case of single-star progenitors is relatively simple, but it becomes much more complicated when including binary-star evolution.
The likely important role of binary-star evolution to the formation of
SN progenitors remains to be thoroughly explored (see e.g., \citealt{2007AA...465L..29C}).
Overall, these complications make the final mass, radius and age of the
CCSN progenitors uncertain \citep{2010MNRAS.408..827D}.
For a general use purpose, the age, mass, effective temperature, luminosity, gravitational acceleration, radius, and star type of progenitors are included in the final database.
Here we show the results that are calculated via BSE code \citep{2002MNRAS.329..897H}, as most works used this code. However, the similar results are also calculated using an updated version of the code \citep{2019MNRAS.485..889S}.

Figs. 1--3 show the distribution of CCSN progenitors in various spaces.
Fig. 1 shows the progenitor distribution in the initial mass versus final mass plane. We observe some difference between single- and binary-star progenitors clearly. The final mass of single-star progenitor is lower than about 12 \dsm{}, but that of binary-star progenitor can be as large as twice as some secondaries accreted masses from their primaries. Note that different stellar evolution codes usually give different final masses. The difference among the results can be as large as 6 \dsm{} (see the comparison of results of, e.g. \citealt{2000ApJ...528..368H,2018PhDT.......126L,2018MNRAS.473.3863L}).

Fig. 2 shows the progenitor distribution in the age versus initial mass plane. We observe that the age of single-star progenitors decreases with increasing initial mass. Meanwhile, the case of binary-star progenitor is much more complicated. In particular, some binary-star progenitors have significantly older ages than those single-star progenitors. The reason is that a long time is needed for the mass exchange of these binary-star progenitors.

Fig. 3 shows the progenitor distribution in the gravity versus effective temperature plane. We see that many primaries of binary-star CCSN progenitors locate in the high-temperature ($T_{\rm eff} > 10^5 K$) area while there are much less secondary components in this region. This implies that primaries contribute more to the combined SEDs at short-wavelengths, which is verified by the example of Fig. 4. That figure gives the contributions of two components of binary-star progenitors to the combined SED. From the figure we also find that both two components of binary-star progenitors contribute to the combined SEDs at long-wavelengths.

When comparing the final masses of CCSN progenitors to the results of \cite{lisakov_tel-02018238} and \cite{2010MNRAS.408..827D},
the results of this work (calculated via BSE) are found to consistent with those calculated via MESA code \citep{2011ApJS..192....3P}. Table 3 shows the results of different works.

\begin{figure*}%1
%  \vspace*{174pt}
 \begin{center}
  \includegraphics[angle=-90,width=\textwidth]{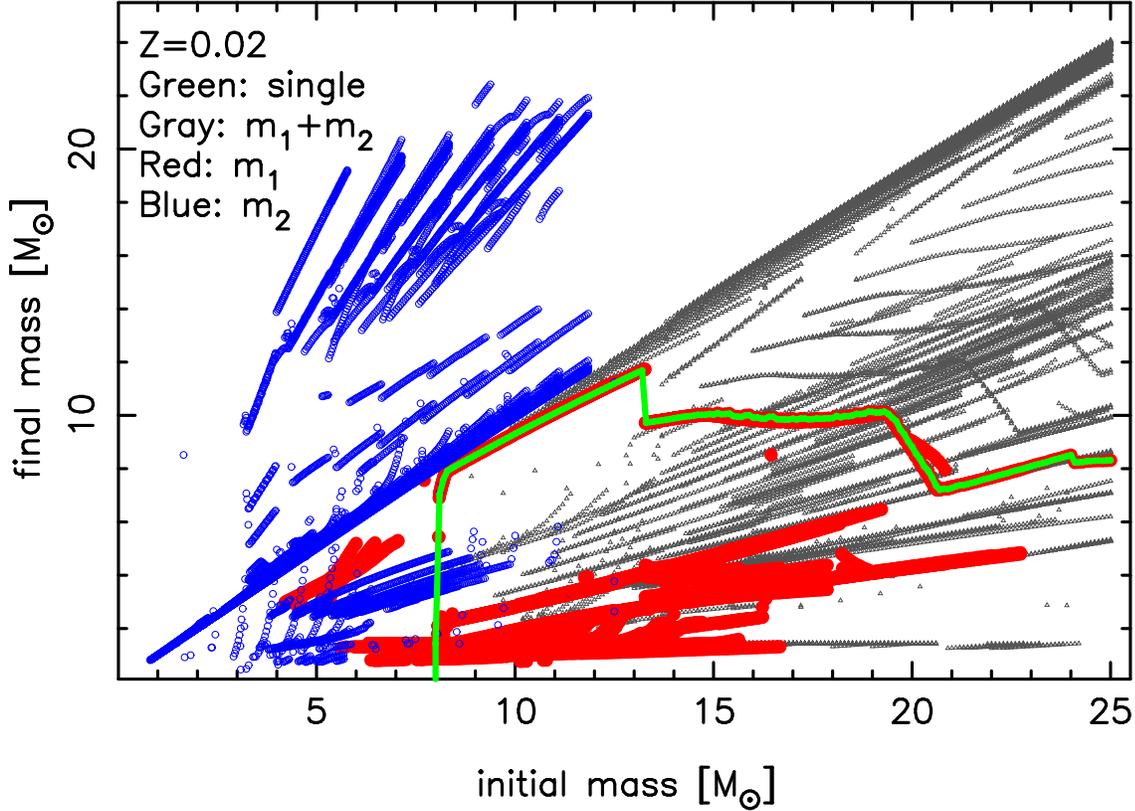}
 \end{center}
 \caption{Final mass as a function of initial mass for solar-metallicity CCSN progenitors.
 Green points are for single-star progenitors, while red points and blue circles for the primary and secondary components of binary-star progenitors.
 Gray triangles are for the total mass of binary-star progenitors.}
\end{figure*}

\begin{figure*}%2
%  \vspace*{174pt}
 \begin{center}
  \includegraphics[angle=-90,width=\textwidth]{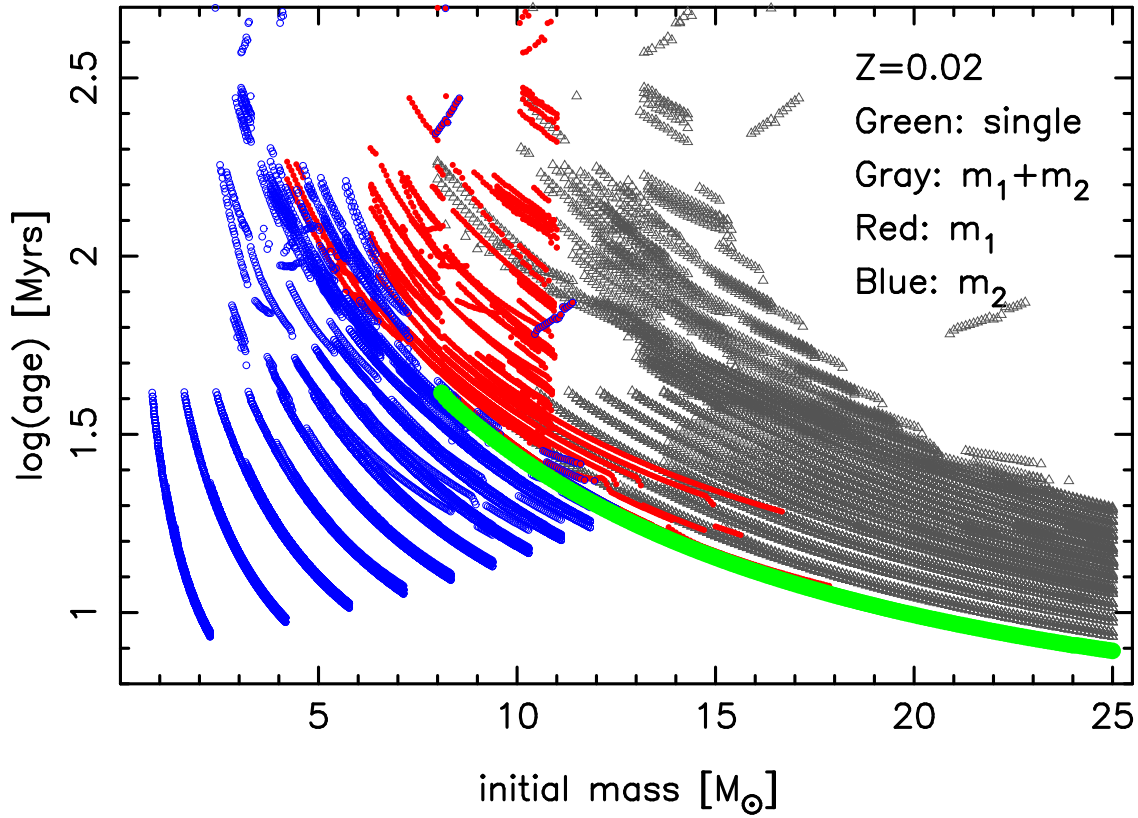}
 \end{center}
 \caption{Age as a function of initial mass for solar-metallicity CCSN progenitors. Colors have the same meanings as in Fig. 1.}
\end{figure*}

\begin{figure*}[htbp]%3
\centering
\begin{minipage}[c]{0.45\textwidth}
\centerline{\includegraphics[angle=-90,scale=.328]{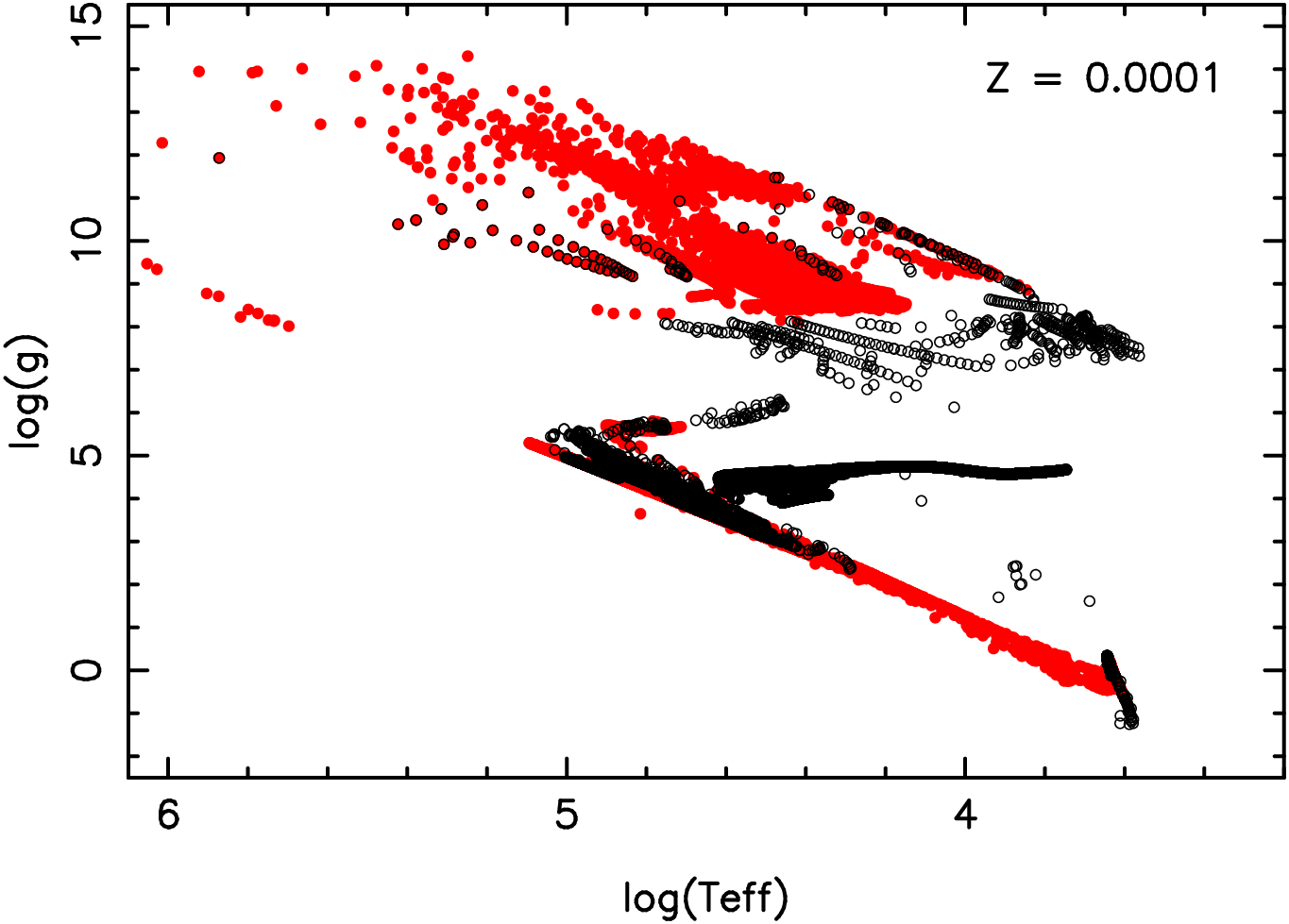}}
\label{ago2pic2}
\end{minipage}
\hfill
\begin{minipage}[c]{0.45\textwidth}
\centering
\centerline{\includegraphics[angle=-90,scale=.35]{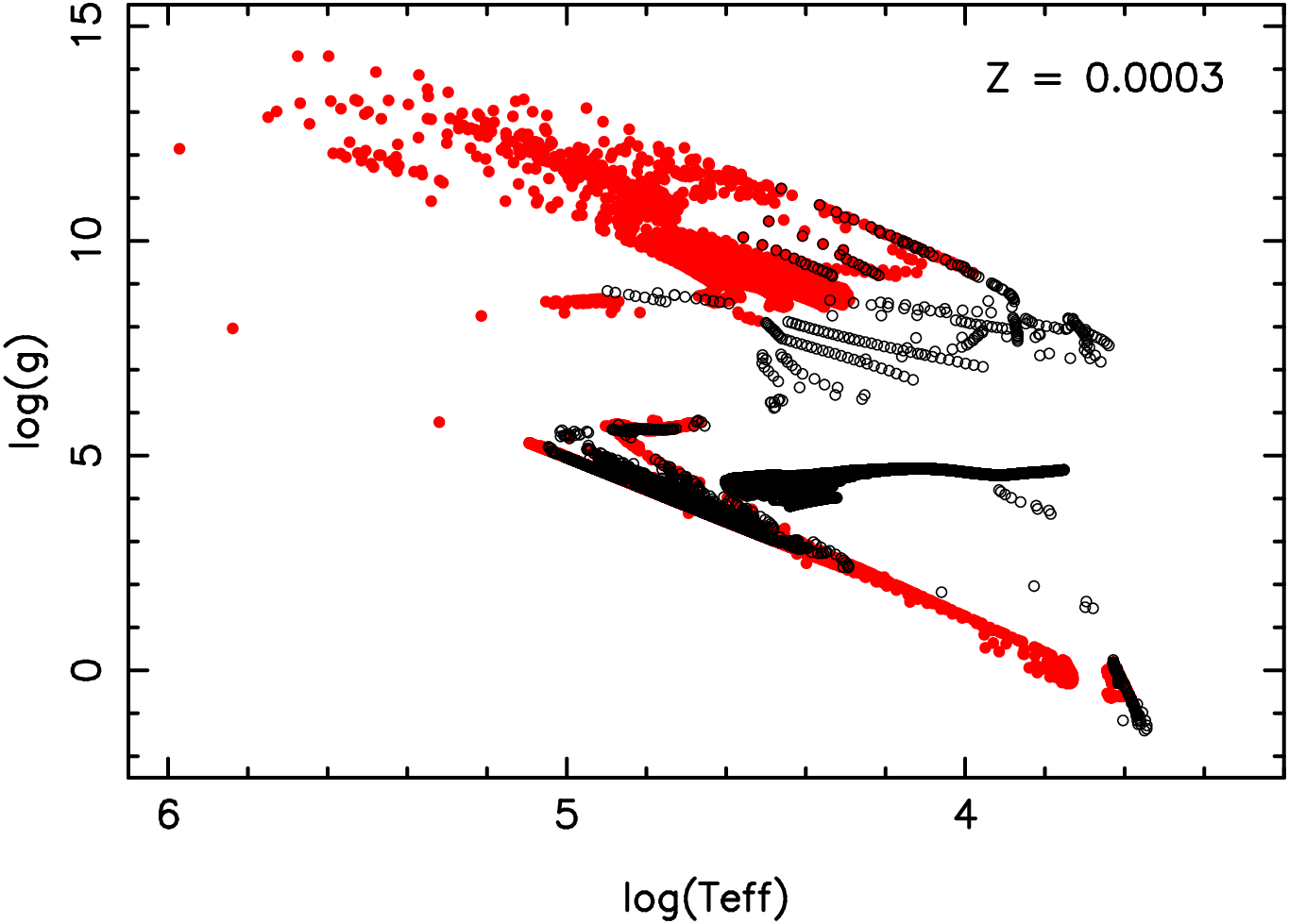}}
\label{ago2pic3}
\end{minipage}
\begin{minipage}[c]{0.45\textwidth}
\centerline{\includegraphics[angle=-90,scale=.328]{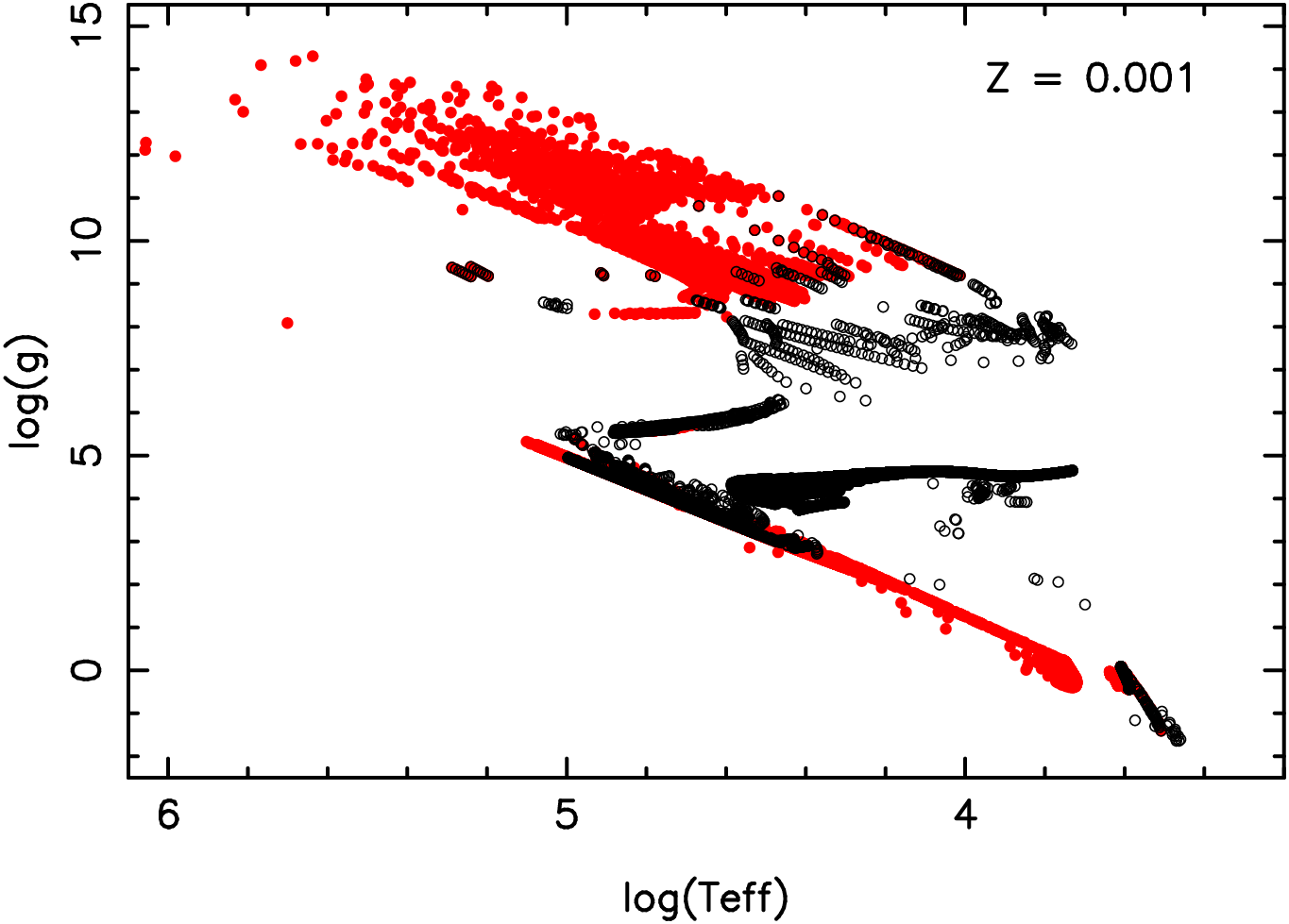}}
\label{ago2pic2}
\end{minipage}
\hfill
\begin{minipage}[c]{0.45\textwidth}
\centering
\centerline{\includegraphics[angle=-90,scale=.35]{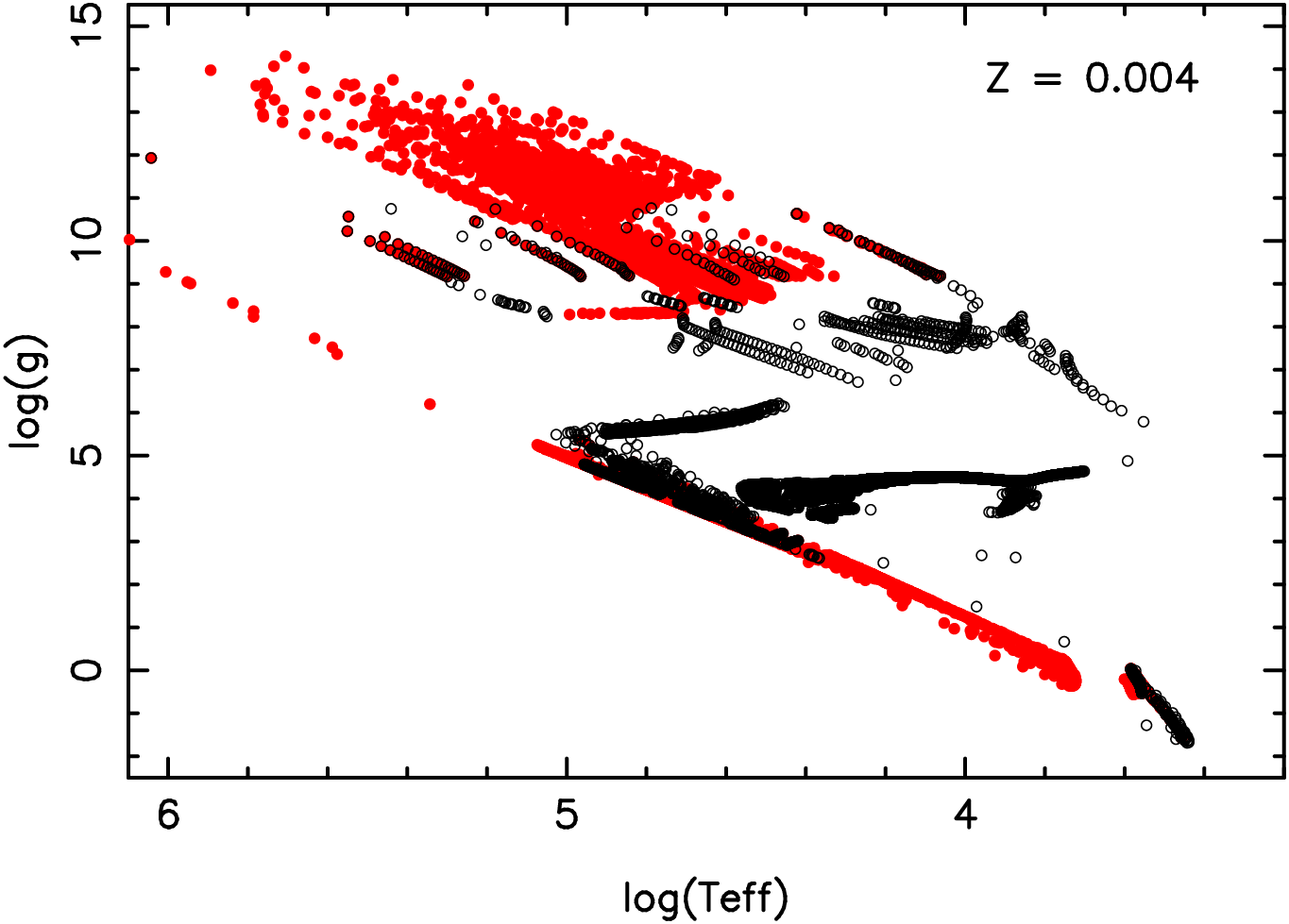}}
\label{ago2pic3}
\end{minipage}
\begin{minipage}[c]{0.45\textwidth}
\centerline{\includegraphics[angle=-90,scale=.328]{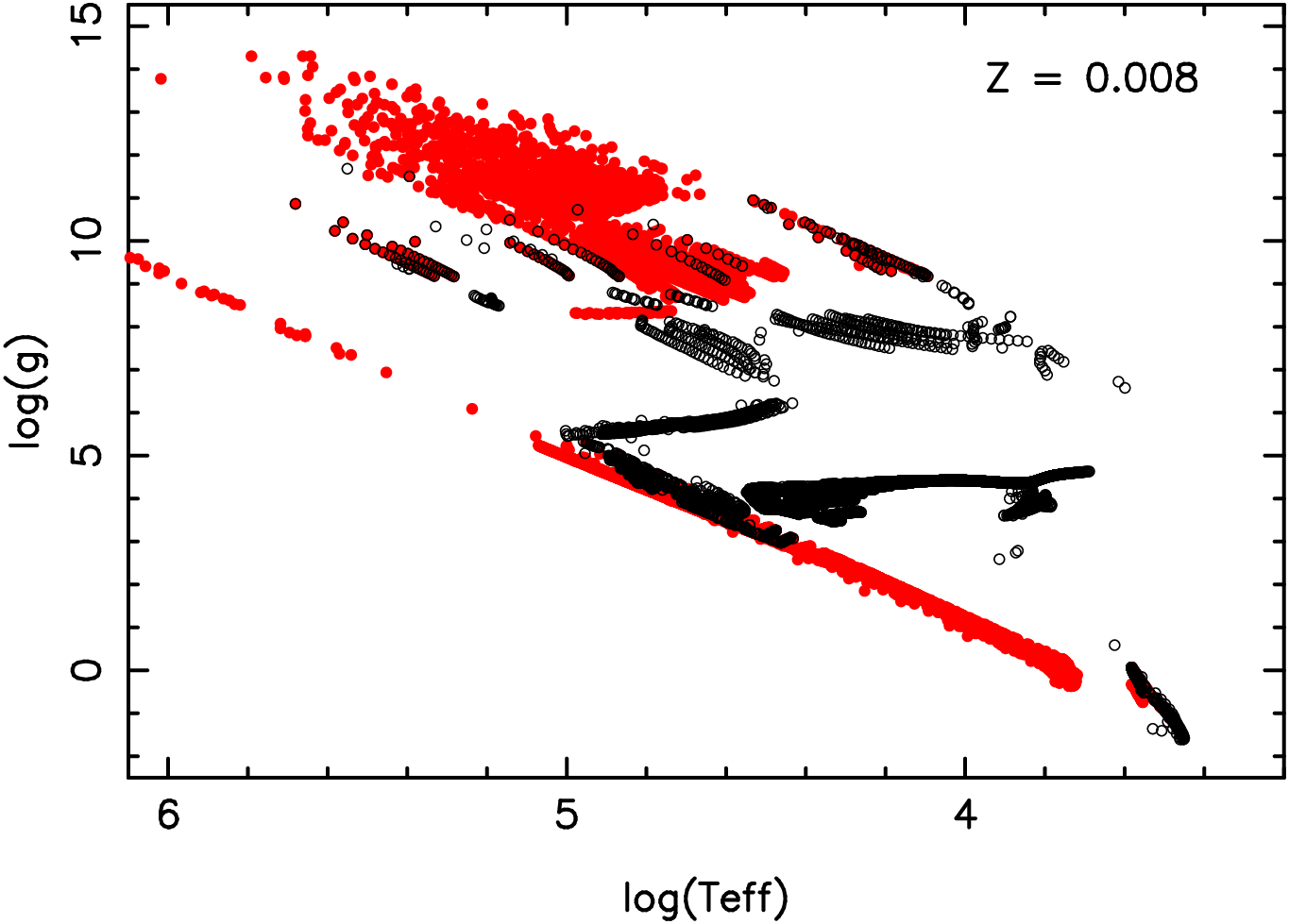}}
\label{ago2pic2}
\end{minipage}
\hfill
\begin{minipage}[c]{0.45\textwidth}
\centering
\centerline{\includegraphics[angle=-90,scale=.35]{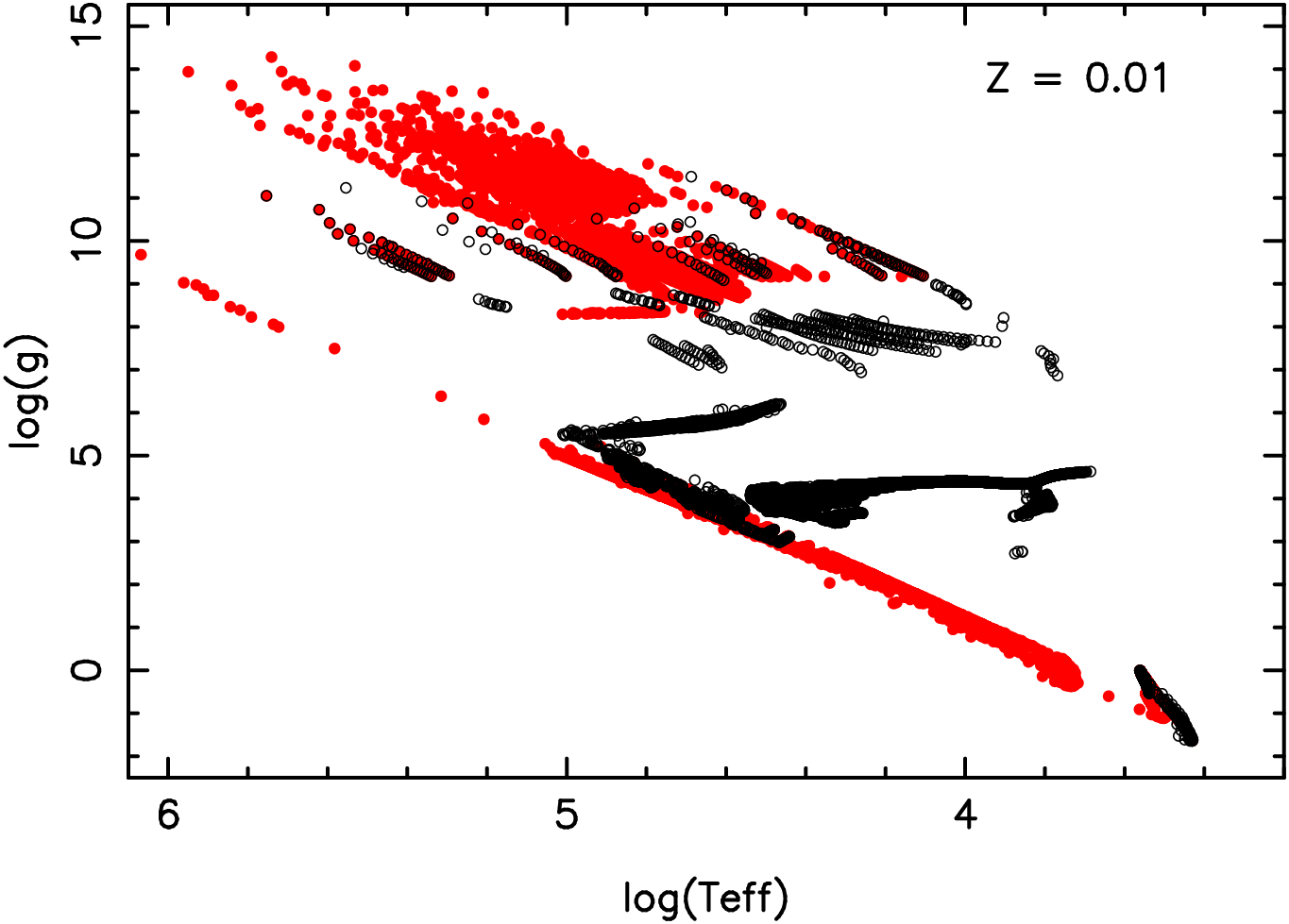}}
\label{ago2pic3}
\end{minipage}
\begin{minipage}[c]{0.45\textwidth}
\centerline{\includegraphics[angle=-90,scale=.328]{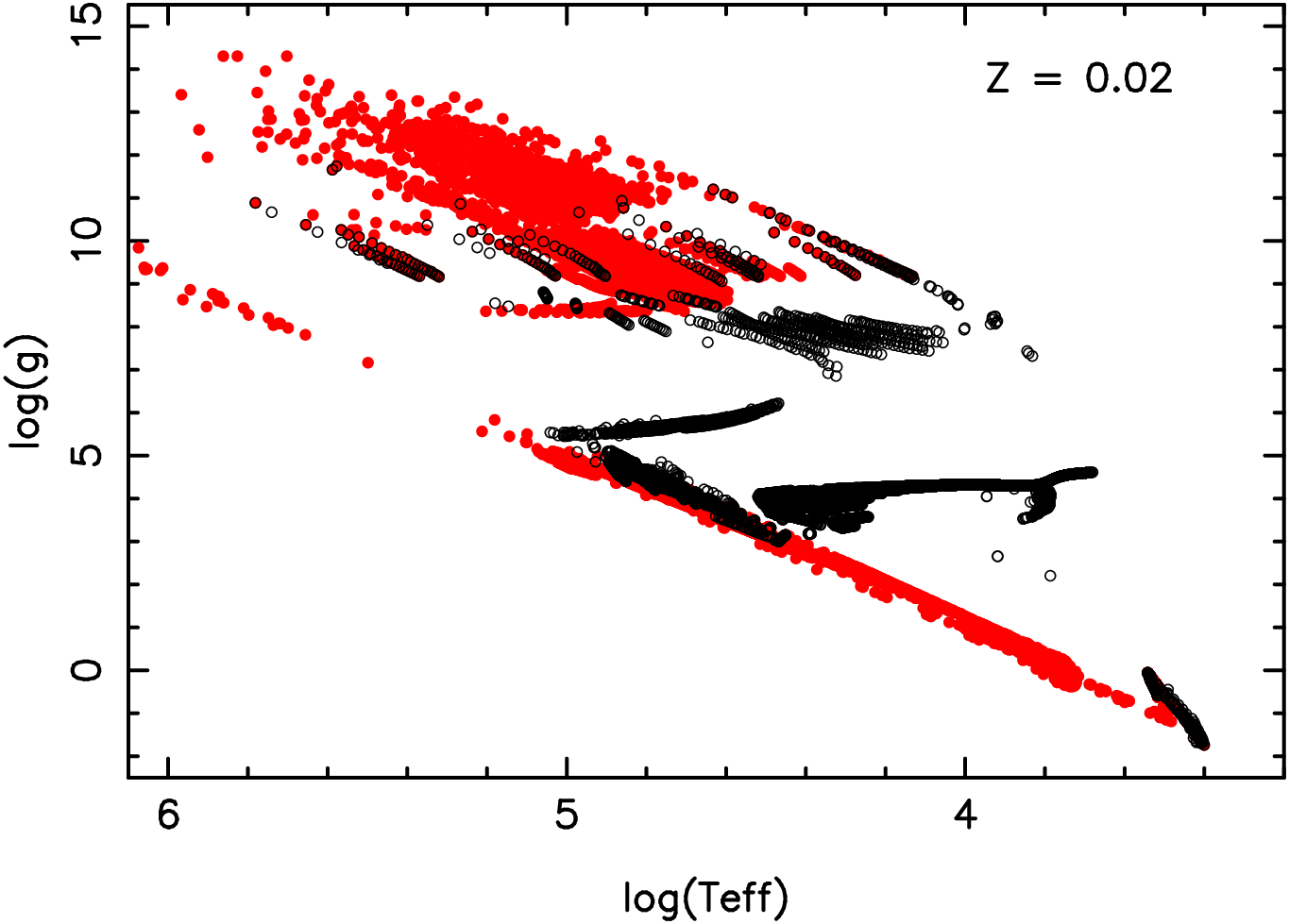}}
\label{ago2pic2}
\end{minipage}
\hfill
\begin{minipage}[c]{0.45\textwidth}
\centering
\centerline{\includegraphics[angle=-90,scale=.35]{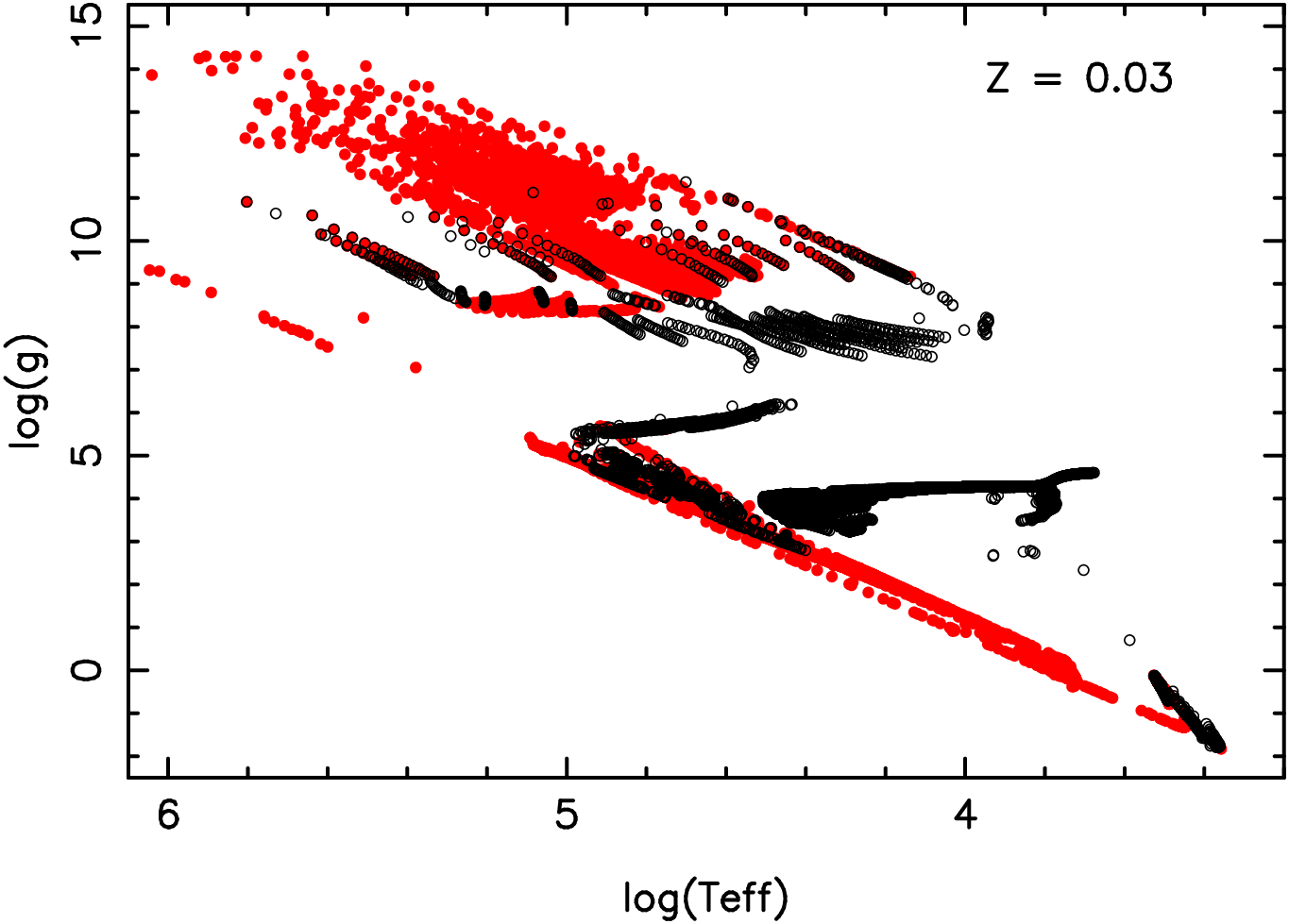}}
\label{ago2pic3}
\end{minipage}
\caption{Distribution of CCSN progenitors in the plane of gravity (log\emph{g}) versus effective temperature ($T_{\rm eff}$). $T_{\rm eff}$ is in \emph{K}. Red filled and black open circles are for the primaries and secondaries of binary-star CCSN progenitors, respectively.}
\end{figure*}

\begin{figure*}%4
%  \vspace*{174pt}
 \begin{center}
  \includegraphics[angle=-90,width=\textwidth]{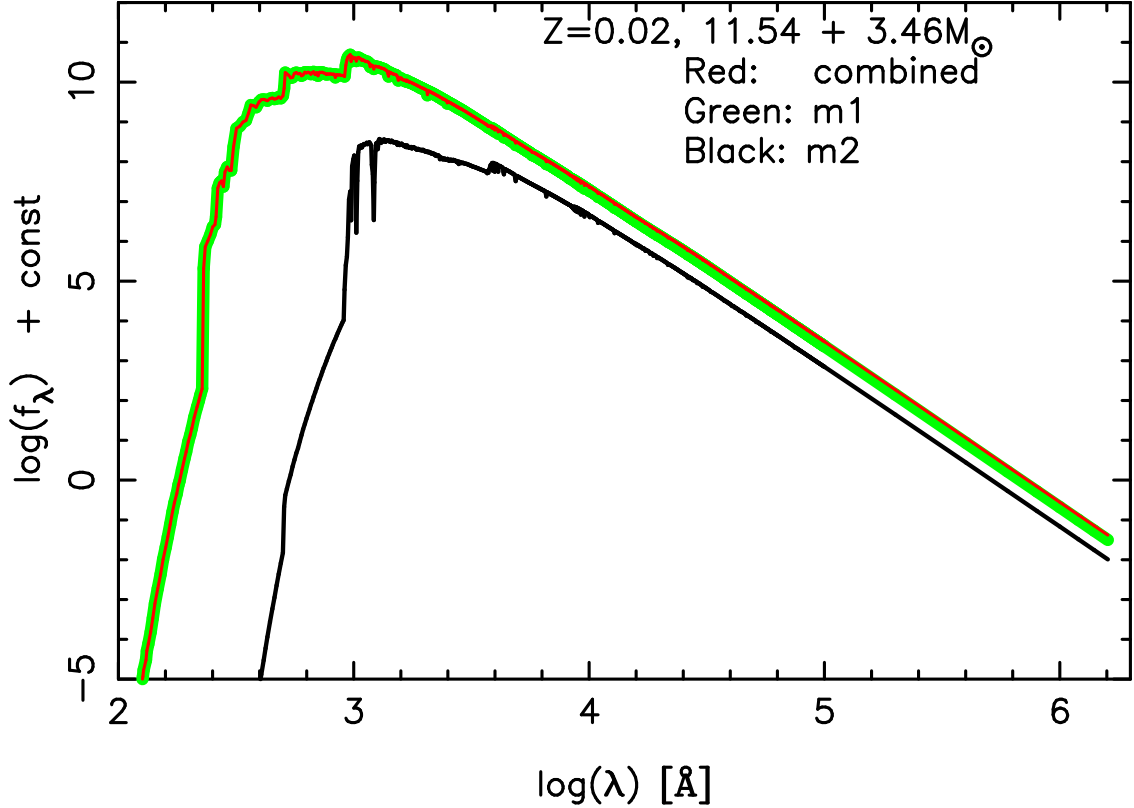}
 \end{center}
 \caption{Contributions of two components to the combined SED of a binary-star CCSN progenitor. Tow components have the solar metallicity, and initial masses of 11.54\dsm{} and 3.46\dsm{}. Green and black lines are for the SEDs of primary and secondary components respectively, while red line is for the combined SED.}
\end{figure*}

\begin{table*} %Table 3
 \caption{Comparison of the final mass and age of CCSN progenitors of this work to two previous works \citep{lisakov_tel-02018238,2010MNRAS.408..827D}. Subscripts `BSE' corresponds to the results of this work, while `MESA' and `Woosely' denote the results that are calculated via MESA \citep{2011ApJS..192....3P} and \cite{2002RvMP...74.1015W} codes, respectively. All models have the metallicity of $Z$ = 0.02. $M_{\rm init}$ is initial main sequence mass.}
 \begin{center}
  \begin{tabular}{ccccccccccc}
   \hline
   $M_{\rm init}$&  $M_{\rm BSE}$&   $M_{\rm MESA}$& $M_{\rm Woosley}$&  Age$_{\rm BSE}$&  Age$_{\rm MESA}$\\
   \hline
   \dsm{}& \dsm{}& \dsm{}&  \dsm{}& Myr& Myr\\
   \hline
  13.0&  11.5&  11.1&&   17.7&   15.3\\
  15.0&  10.0&  11.9&   12.64&   14.3&   12.5\\
  17.0&   9.9&  14.2&&   12.0&   10.7\\
  19.0&  10.1&  13.6&&   10.5&    9.4\\
  21.0&   7.3&   8.6&&    9.3&    8.5\\
  23.0&   8.1&   8.1&&    8.5&    7.7\\
  25.0&   8.3&   8.6&  12.53&    7.8&    7.2\\
   \hline
  \end{tabular}
 \end{center}
\end{table*}

\section{Spec-SEDs}
This section presents the spec-SEDs of both single- and binary-star CCSN progenitors.
Because the database is as large as 1.4\,GB, only some examples are shown here.
One can see Figs. 5--7 for the spec-SEDs of a few example progenitors with metallicities of 0.001, 0.02 (solar metallicity) and 0.03, and total main-sequence masses of 9.1 \dsm{} (solid lines) and 20 \dsm{} (dashed lines).
We read that there is obvious difference between the SEDs of single- and binary-star progenitors, even though they have the same metallicity and total mass.
This suggests that different results will be possibly obtained when fitting to the observed SEDs using single- and binary-star progenitor models.

\begin{figure*}%5
%  \vspace*{174pt}
 \begin{center}
  \includegraphics[angle=-90,width=\textwidth]{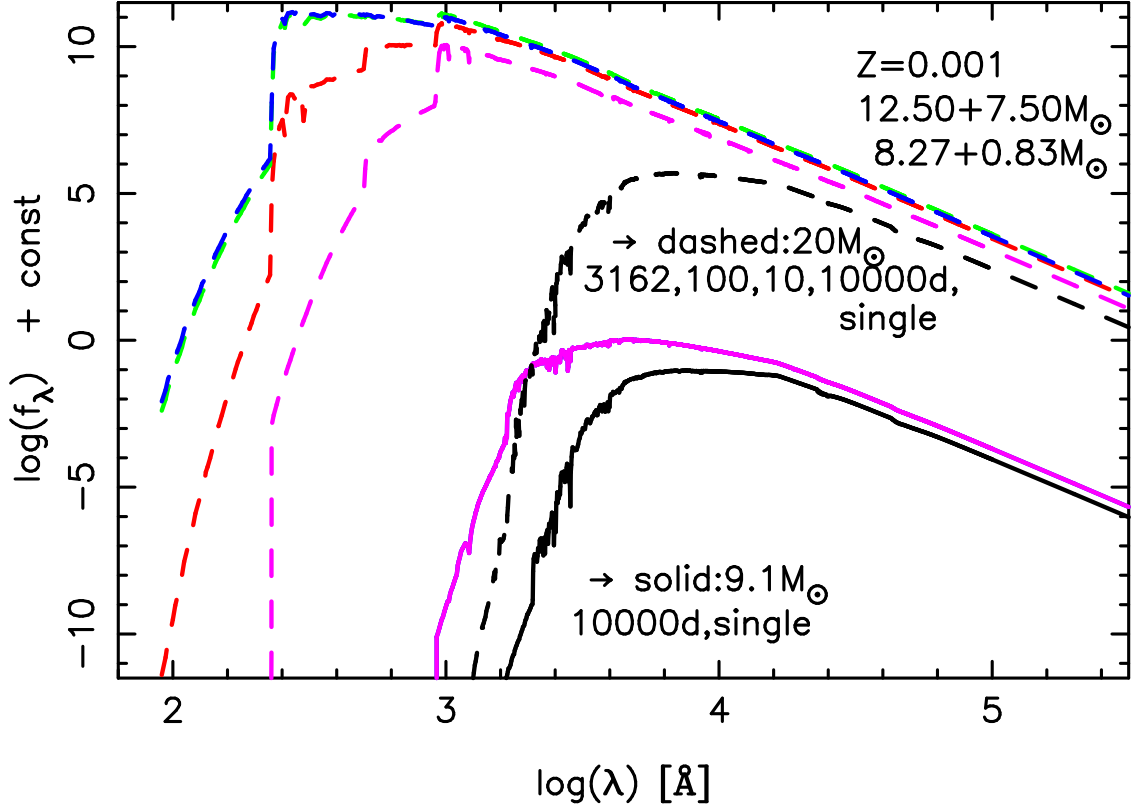}
 \end{center}
 \caption{Spec-SEDs of example CCSN progenitors. The metallicity $Z$ is 0.001. Solid and dashed lines are for 9.1\dsm{} and 20\dsm{} models respectively. ``single'' denotes single-star progenitors, while orbital period numbers denote binary-star progenitors. Blue, green, red, and purple lines are for orbital periods of 3162, 100, 10, and 10000\,d, while black line is for single-star progenitors.}
\end{figure*}

\begin{figure*}%6
%  \vspace*{174pt}
 \begin{center}
  \includegraphics[angle=-90,width=\textwidth]{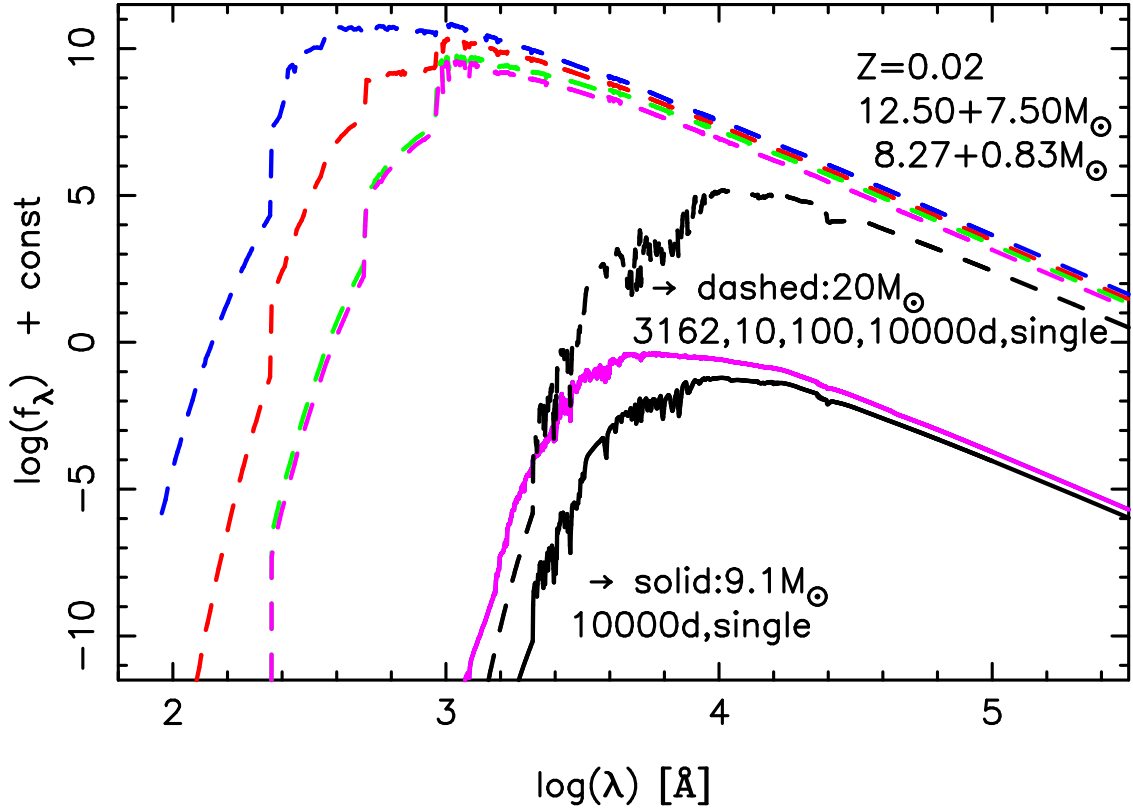}
 \end{center}
 \caption{Similar to Fig. 5, but for a solar metallicity of $Z$ = 0.02.}
\end{figure*}

\begin{figure*}%7
%  \vspace*{174pt}
 \begin{center}
  \includegraphics[angle=-90,width=\textwidth]{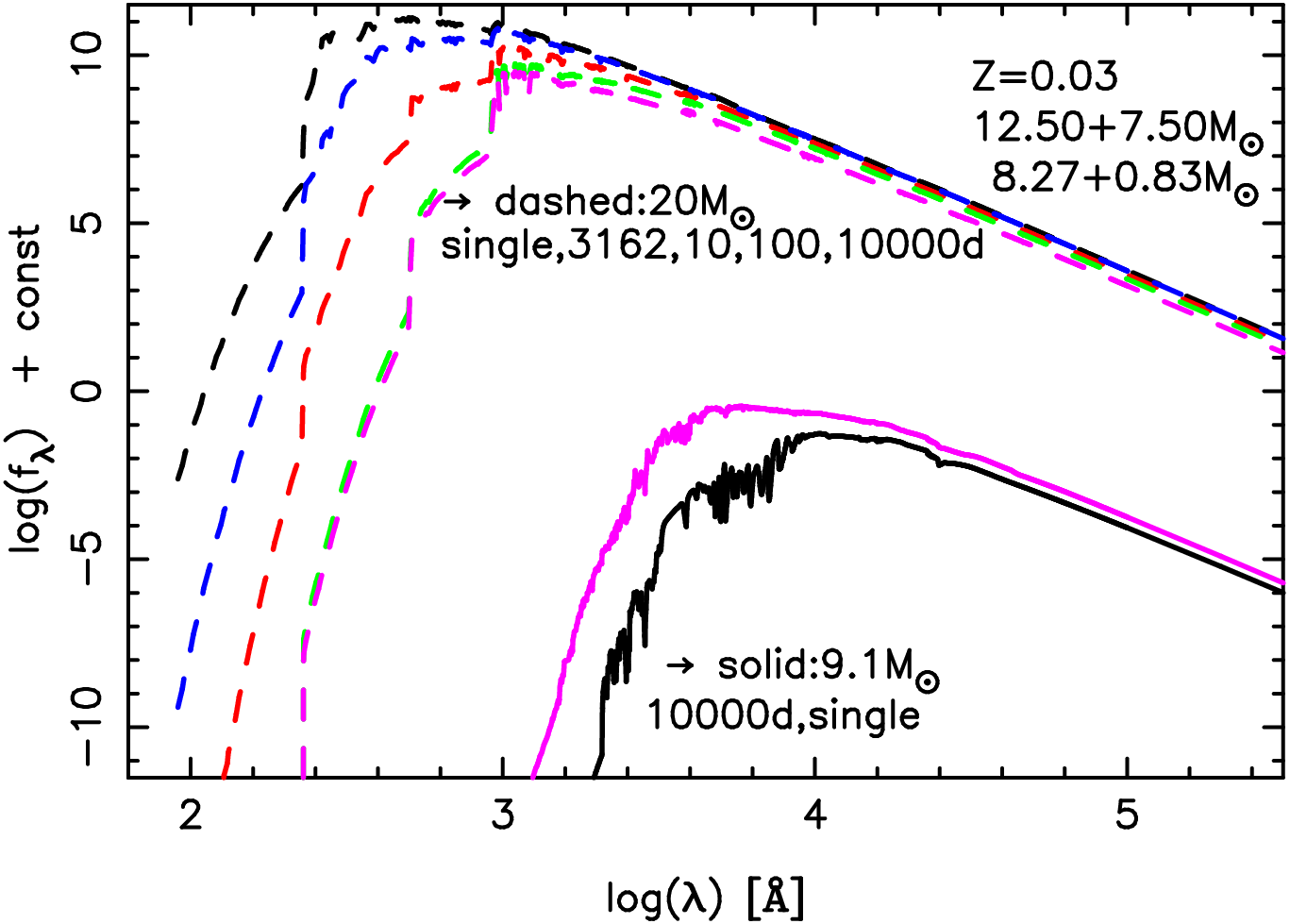}
 \end{center}
 \caption{Similar to Fig. 5, but for a metallicity of $Z$ = 0.03.}
\end{figure*}

\section{Phot-SEDs}
This section shows the phot-SEDs of CCSN progenitors. Such SEDs are usually more useful for the studies of CCSN progenitors.
All phot-SEDs are calculated from the spec-SEDs.
The AB system are adopted, for a purpose of wide applications. As a result, the AB magnitudes in $FUV$, $NUV$, $u$, $g$, $r$, $i$, $z$, $J$, $H$, $Ks$, $W1$, $W2$, and $W3$ bands are calculated. This makes it possible to study the phot-SEDs of CCSN progenitors in a wide wavelength range. Figs. 8--10 show some examples of the phot-SEDs of single- and binary-star progenitors. Fig. 8 shows the phot-SEDs of single-star progenitors, for eight metallicities from 0.0001 to 0.03. It is shown that single-star progenitors with various masses and metallicities usually have different phot-SED shapes. Some massive progenitors with metallicity poorer than 0.001 have UV-upturn phot-SEDs.
However, there is obvious overlap for the phot-SEDs of single-star progenitors. This implies that the metllicity and main-sequence mass of such progenitors can be determined via fitting to the observed SEDs, but the uncertainties of the results of some progenitors will be possibly large. This agrees with previous studies on SNe such as SN1987A. Similarly, Figs. 9 and 10 show the phot-SEDs of some example binary-star CCSN progenitors.

\begin{figure*}%8
%  \vspace*{174pt}
 \begin{center}
  \includegraphics[angle=-90,width=\textwidth]{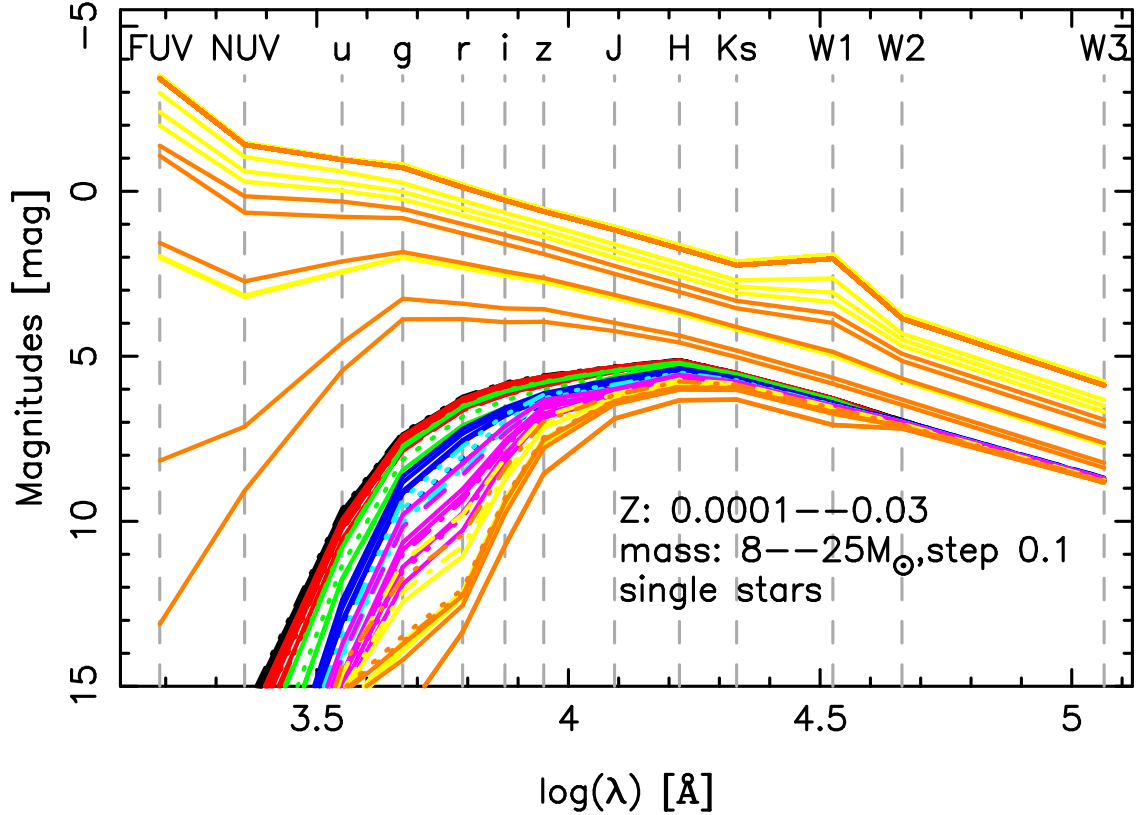}
 \end{center}
 \caption{Example phot-SEDs of single-star CCSN progenitors. Black, red, green, blue, cyan, purple, yellow, and orange colors are for $Z$ = 0.0001, 0.0003, 0.001, 0.004, 0.008, 0.01, 0.02, 0.03, respectively. Lines with the same color but different shapes are for various masses.}
\end{figure*}

\begin{figure*}%9
%  \vspace*{174pt}
 \begin{center}
  \includegraphics[angle=-90,width=\textwidth]{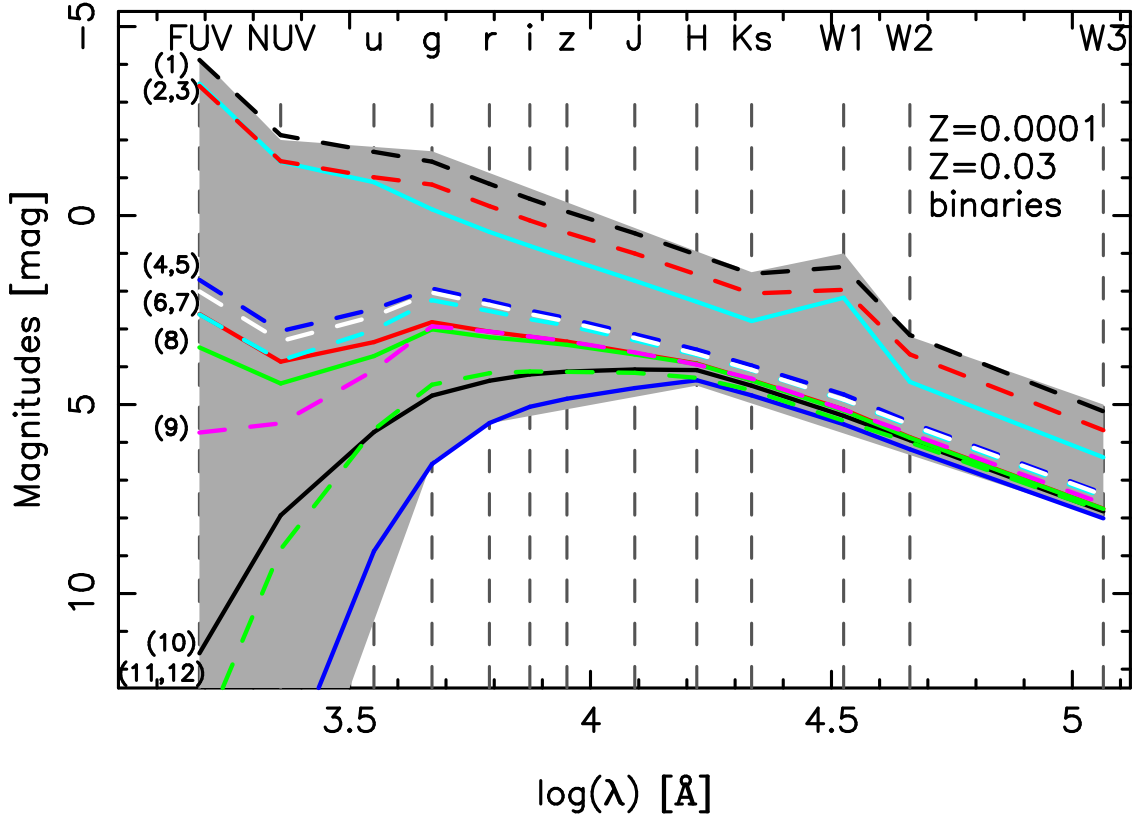}
 \end{center}
 \caption{Example phot-SEDs of binary-star CCSN progenitors with metallicities of 0.0001 and 0.03. Gray area indicates the range of all phot-SEDs of all binary-star progenitors in the database. The detailed model parameters of these CCSN progenitors are listed in Table 4.}
\end{figure*}

\begin{figure*}%10
%  \vspace*{174pt}
 \begin{center}
  \includegraphics[angle=-90,width=\textwidth]{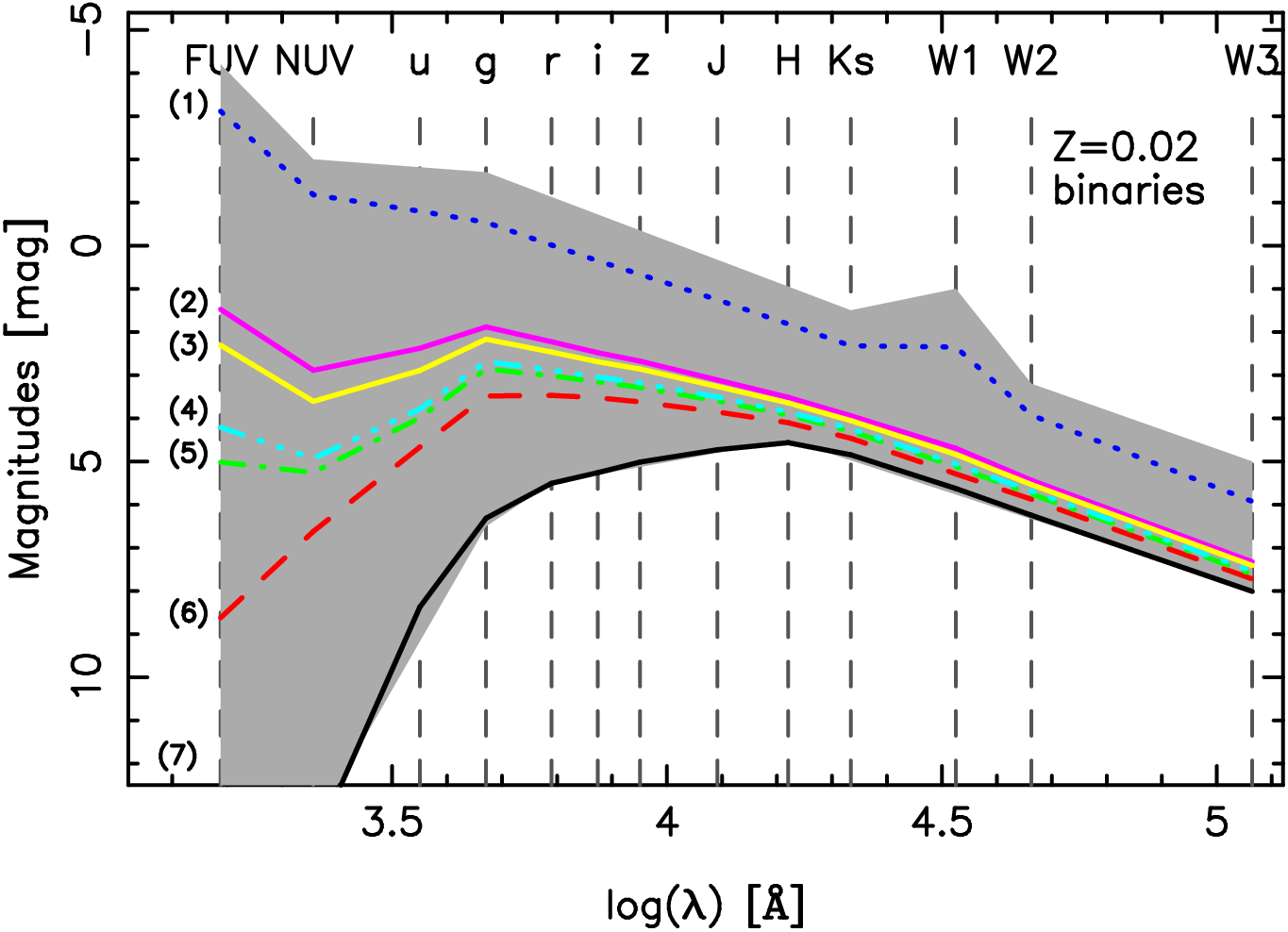}
 \end{center}
 \caption{Similar to Fig. 9, but for a metallicity of $Z$ = 0.02.}
\end{figure*}

\begin{table*} %Table 4
 \caption{CCSN progenitor models for Figs. 9 and 10. ``No.'' means the line number in two figures. $m_{\rm 1}$ and $m_{\rm 2}$ are in \dsm{}, and $P$ is in days.}
 \begin{center}
  \begin{tabular}{ccccccccccccc}
   \hline
   No.     &$m_{\rm 1}$   &$m_{\rm 2}$   &$P$     &$e$&&No.     &$m_{\rm 1}$   &$m_{\rm 2}$   &$P$     &$e$\\
   \hline
   &&Fig. 9&&  &&&&Fig. 10&&\\
   \hline
    1&      8.18&      0.82&   3162&      0.3&&   1&      8.18&      0.82&  10000&      0.3\\
    2&      8.18&      0.82&  10000&      0.3&&   2&     16.36&      1.64&  10000&      0.3\\
    3&      7.50&      1.50&   1000&      0.3&&   3&     19.09&      1.91&   3162&      0.3\\
    4&     10.91&      1.09&   3162&      0.3&&   4&     17.50&      3.50&    316&      0.3\\
    5&     13.64&      1.36&   3162&      0.3&&   5&     20.00&      2.00&   3162&      0.3\\
    6&     16.36&      1.64&      3&      0.7&&   6&      8.57&      3.43&   3162&      0.3\\
    7&     16.36&      1.64&   3162&      0.3&&   7&      9.23&      2.77&   3162&      0.3\\
    8&     10.91&      1.09&  10000&      0.3\\
    9&     10.00&      2.00&   3162&      0.3\\
    10&     9.23&      2.77&   3162&      0.3\\
    11&    13.64&      1.36&  10000&      0.3\\
    12&    12.50&      2.50&  10000&      0.3\\
   \hline
  \end{tabular}
 \end{center}
\end{table*}

\section{Application of SED database to mock progenitors}
This section applies the database to some mock CCSN progenitors.
The phot-SEDs of mock progenitors are fitted using the database.
Each phot-SED consists of the magnitudes in $FUV$, $NUV$, $u$, $g$, $r$, $i$, $z$, $J$, $H$, $Ks$, $W1$, $W2$, and $W3$ bands.
It is found that the main-sequence mass, age and metallicity of most single-star progenitors can be reproduced as a whole,
although a few progenitors are not reproduced well because of the metallicity and mass degeneracy.
As an example, Fig. 11 shows the comparison of input and reproduced masses of single-star progenitors.

However, the main-sequence masses of most binary-star progenitors are not reproduced correctly, if all parameters are free in the SED fitting (Fig. 12).
The fitted main-sequence masses of most binary-star progenitors are much lower than the real values.
This is caused by the degeneracy among mass, metallicity and orbital period.
In order to find a reliable method for determining the main-sequence masses of binary-star progenitors,
the cases of fixed metallicity or fixed period are tested, but the uncertainties in results are still large.
Finally, the case of fixed metallicity and period gives satisfactory results (see Fig. 13).
This means that if one wants to determine the masses of bianry-star progenitors reliably,
the metallicity and orbital period (initial or final values) are suggested to be determined in advance.

\begin{figure*}%11
%  \vspace*{174pt}
 \begin{center}
  \includegraphics[angle=-90,width=\textwidth]{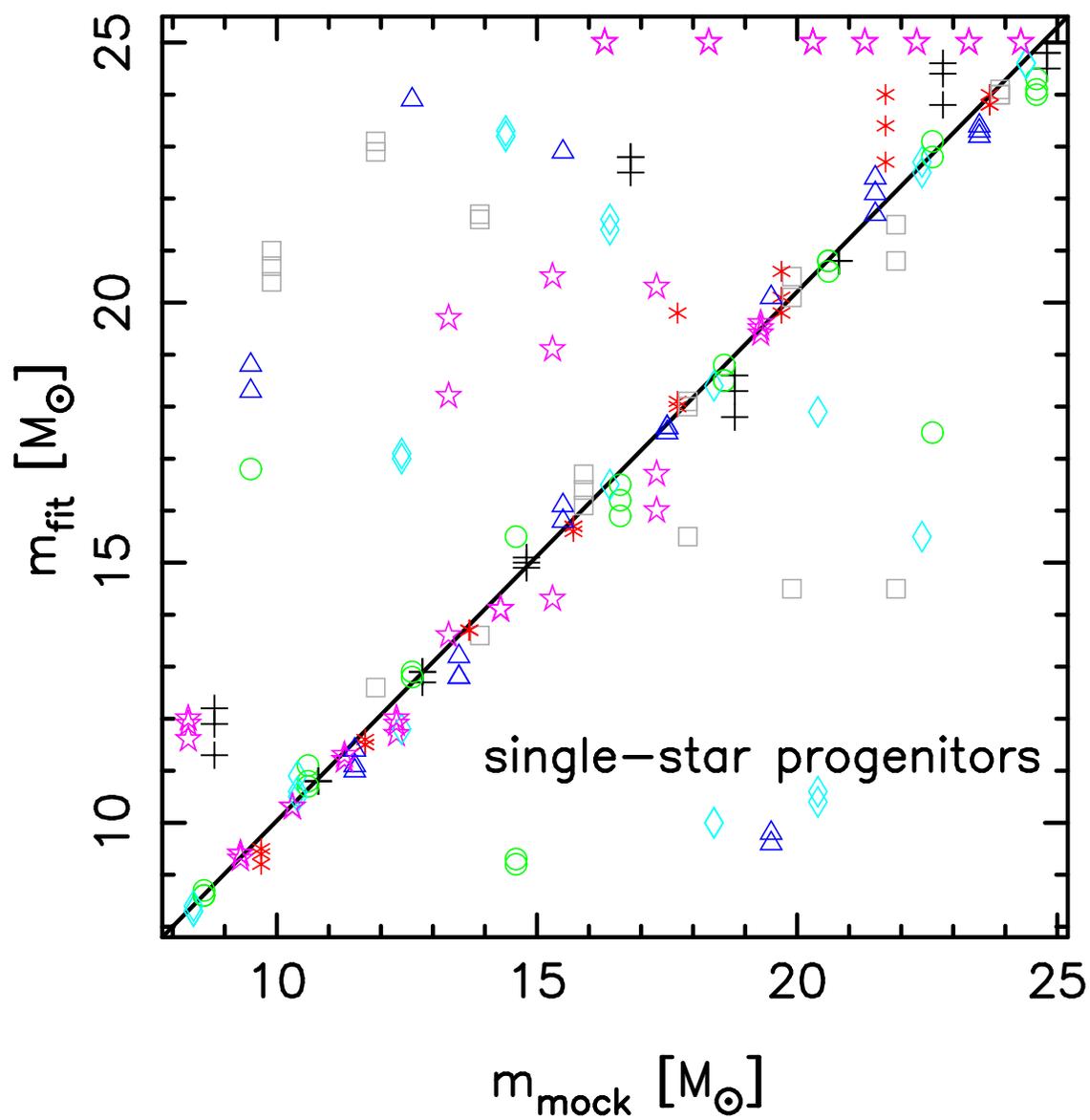}
 \end{center}
 \caption{Comparison of input ($m_{\rm mock}$) and reproduced ($m_{\rm fit}$) masses of single-star progenitors in phot-SED fitting. Different symbols denote different metallicities.}
\end{figure*}

\begin{figure*}%12
%  \vspace*{174pt}
 \begin{center}
  \includegraphics[angle=-90,width=\textwidth]{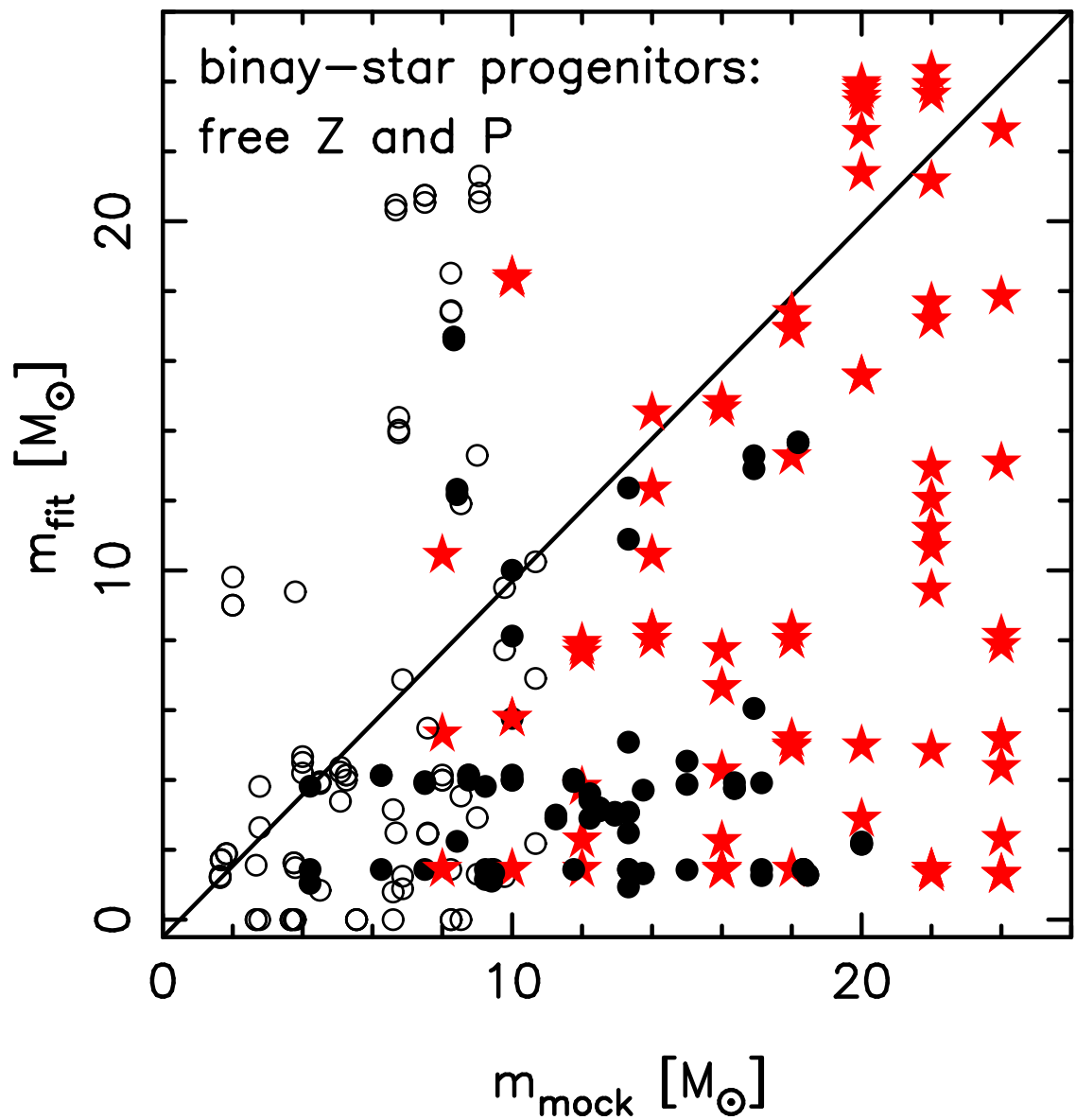}
 \end{center}
 \caption{Comparison of input ($m_{\rm mock}$) and reproduced ($m_{\rm fit}$) masses of binary-star progenitors in phot-SED fitting. The result is for the case of free metallicity and orbital period. Filled circle, open circle and pentagram are for primary mass, secondary mass and total mass respectively.}
\end{figure*}

\begin{figure*}%13
%  \vspace*{174pt}
 \begin{center}
  \includegraphics[angle=-90,width=\textwidth]{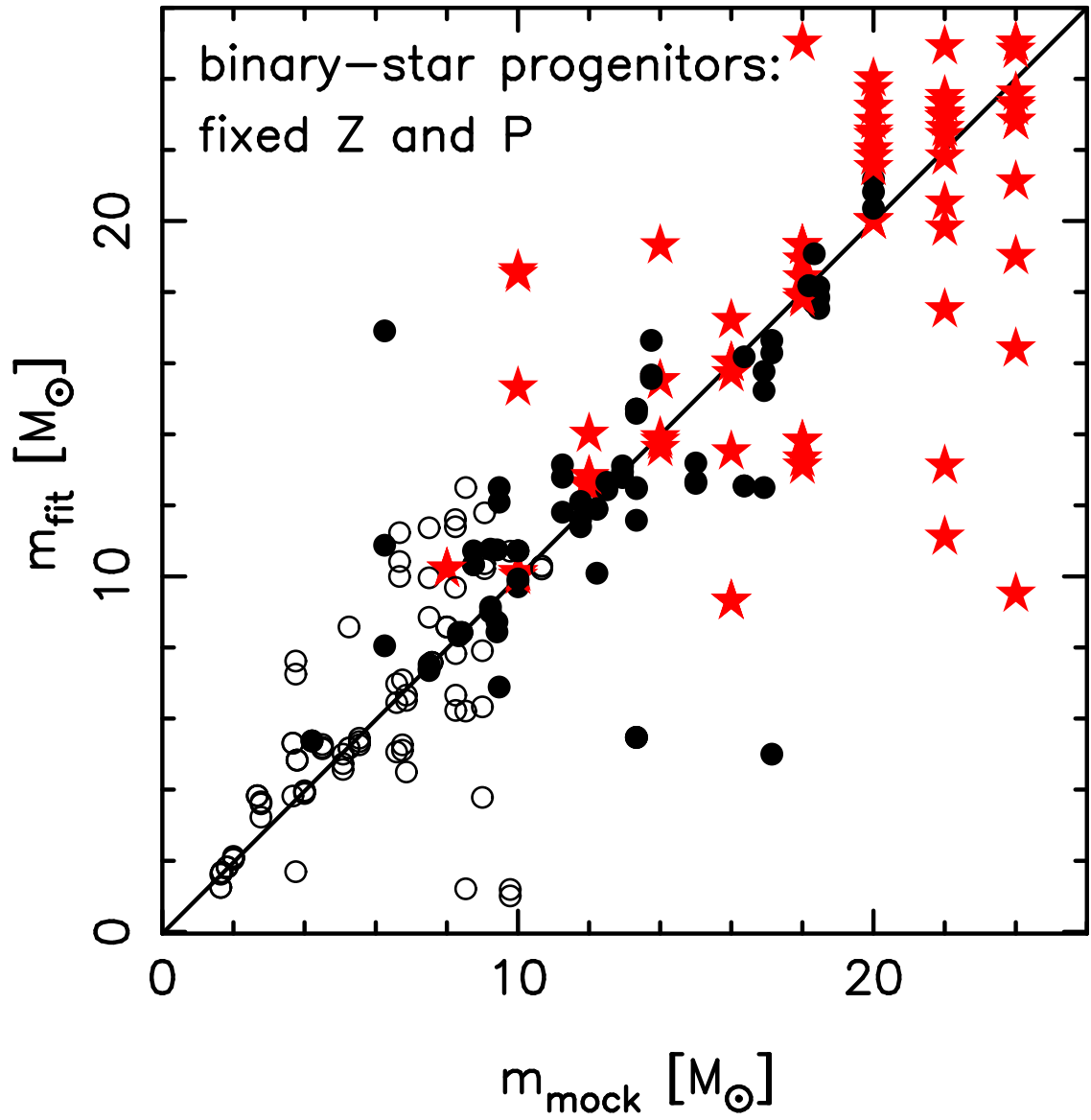}
 \end{center}
 \caption{Similar to Fig. 11, but for binary-star progenitors with known metallicity and orbital period. }
\end{figure*}

\section{Conclusion}

This paper presents a new database of SEDs of the single- and binary-star CCSN progenitors.
Both the phot- and spec-SEDs of progenitors are included in the database.
The database covers wide ranges of metallicity (0.0001--0.03), main-sequence mass (8--25 \dsm{}), component mass ratio (0--1), binary period (0.1--10$^{4}$\,days), and two eccentricities (0.3 and 0.7). It is then applied to the phot-SEDs of some mock CCSN progenitors.
Our investigation leads to the following conclusions:
\begin{itemize}
	 \item The database of spec- and phot-SEDs of CCSN progenitors can be used for the studies of progenitor properties, and the difference between binary- and single-star progenitors. The results are consistent with those calculated via MESA code \citep{2011ApJS..192....3P}, but the database is model dependent. Thus it is better for statistical studies such as population synthesis. It can be potentially used for the identification of CCSN progenitors in large surveys. Stellar evolutionary code can affect the results, but the relative results are usually similar. For example, when we use the code of \cite{2019MNRAS.485..889S} to calculate the SEDs instead of \cite{2002MNRAS.329..897H}, similar results are shown (see Fig. 14).
  \item Binary-star CCSN progenitors have much more complicated parameter spaces than single-star progenitors. It leads to much larger uncertainties in the determination of progenitor properties including component masses, total mass, metallicity and period.
  \item Binaries with component masses less massive than 8 \dsm{} can form CCSN progenitors, although single stars less massive than this value cannot lead to CCSN.
  \item When the SED database is used for determining the properties of CCSN progenitors, whether the progenitor is single or binary star affects the result accuracy significantly. If progenitors are single stars, the initial and final mass can be determined well for most progenitors via phot-SEDs from FUV to W3 bands. However, the results will be not reliable for binary-star progenitors, if metallicity, mass, and period are set as free parameters of fit. In order to get reliable results, the metallicity and binary period (initial or final periods) are needed to be measured using other methods. If these two parameters are known, the masses of binary components can be determined well via SED fitting.
\end{itemize}

\begin{figure*}%14
%  \vspace*{174pt}
 \begin{center}
  \includegraphics[angle=-90,width=\textwidth]{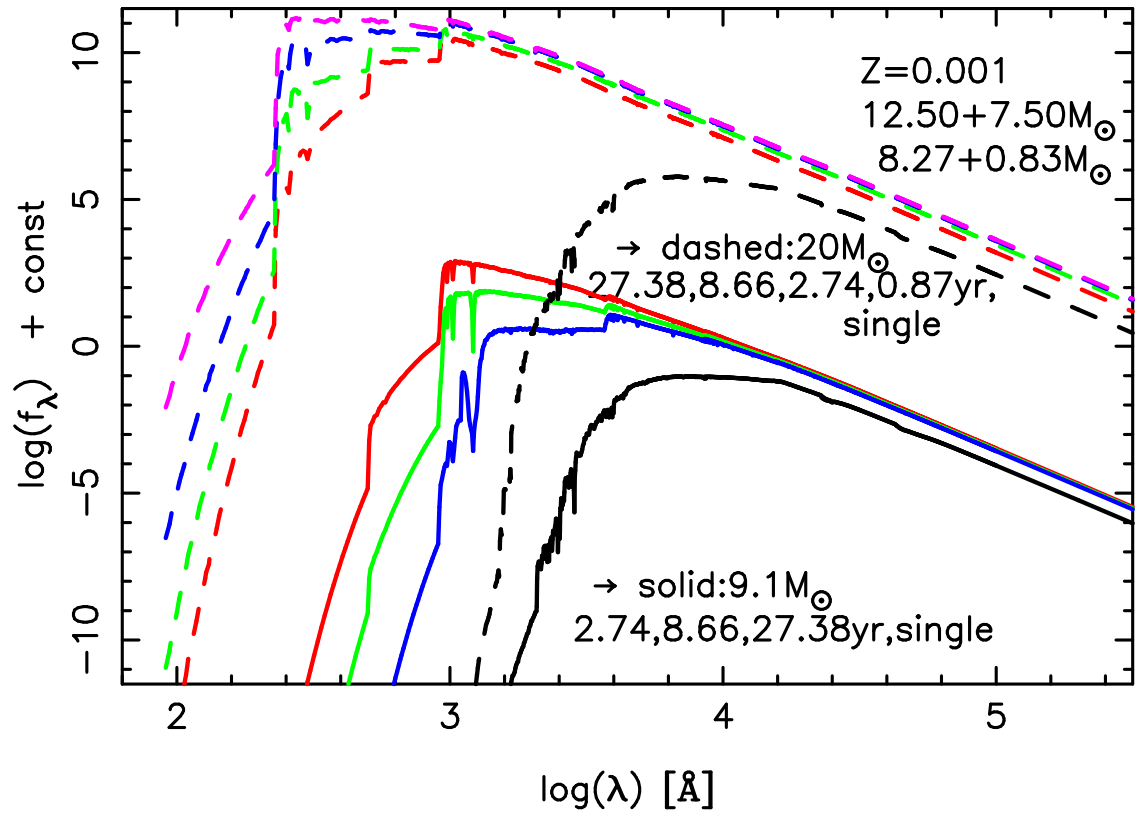}
 \end{center}
 \caption{Similar to Fig. 5, but for another stellar evolution calculation \citep{2019MNRAS.485..889S}. Solid and dashed lines are for total stellar masses of 9.1 and 20\dsm{} respectively.}
\end{figure*}

\begin{acknowledgements}
The authors thank Prof. Xiaofeng Wang for suggestions and Dr. Jicheng Zhang for discussions.
This work has been supported by the Chinese National Science Foundation (No. 11863002),
Sino-German Cooperation Project (No. GZ 1284), and Yunnan Academician Workstation of Wang Jingxiu (No. 201905F150106).
\end{acknowledgements}

\label{lastpage}

\end{document}